\documentclass[pra,10pt,aps]{revtex4-2}
\usepackage{amsmath}
\usepackage{hyperref}
\usepackage{cleveref}
\usepackage{etoolbox}
\usepackage{placeins}
\usepackage{graphicx}
\usepackage{subfig}
\usepackage{makecell}
\usepackage[export]{adjustbox}
\usepackage{tikz}
\usepackage{siunitx}
\usepackage{tabularray}
\usepackage{calc}
\usepackage{mathtools}

\DeclareSIUnit \hbarkU {\ensuremath{\mathit{\hbar k}}}
\newcommand{\hbark}[1]{\qty{#1}{\hbarkU}}

\hypersetup{
    colorlinks=true,
    linkcolor={black},
    anchorcolor={black},
    citecolor={black},
    filecolor={black},
    menucolor={black},
    runcolor={black},
    urlcolor={blue},
    pdfborder={0 0 0},
}

\Crefname{figure}{Figure}{Figures}
\Crefname{equation}{Equation}{Equations}
\Crefname{section}{Section}{Sections}

\crefname{figure}{Fig.}{Figs.}
\crefname{equation}{Eq.}{Eqs.}
\crefname{section}{Section}{Sections}

\newcommand{\phiI}[0]{\theta(i)}
\newcommand{\phiIInp}[0]{\theta_\mathrm{inp}(i)}
\newcommand{\phiIInpD}[0]{\theta_\mathrm{inp}'(i)}
\newcommand{\phiIRec}[0]{\theta_\mathrm{rec}(i)}
\newcommand{\phiIDiff}[0]{\theta_\mathrm{diff}(i)}
\newcommand{\phiXY}[0]{\gamma(x,y)}
\newcommand{\phiXYInp}[0]{\gamma_\mathrm{inp}(x,y)}
\newcommand{\phiXYInpD}[0]{\gamma_\mathrm{inp}'(x,y)}
\newcommand{\phiXYRec}[0]{\gamma_\mathrm{rec}(x,y)}
\newcommand{\phiXYDiff}[0]{\gamma_\mathrm{diff}(x,y)}
\newcommand{\phiIXY}[0]{\Phi(i,x,y)}
\newcommand{\phiIXYInp}[0]{\Phi_\mathrm{inp}(i,x,y)}
\newcommand{\phiIXYInpD}[0]{\Phi'_\mathrm{inp}(i,x,y)}
\newcommand{\portOffset}[0]{\delta x}
\newcommand{\phiIXYLInp}[0]{\Phi_\mathrm{inp}(i,x+\portOffset,y)}
\newcommand{\phiIXYRInp}[0]{\Phi_\mathrm{inp}(i,x-\portOffset,y)}
\newcommand{\phiIXYRec}[0]{\Phi_\mathrm{rec}(i,x,y)}
\newcommand{\phiIStd}[0]{\sigma_i}
\newcommand{\phiIStdMax}[0]{\mathrm{max}(\sigma_i)}
\newcommand{\phiXYStd}[0]{\sigma_{xy}}
\newcommand{\I}[1][i]{I(#1,x,y)}
\newcommand{\A}[0]{A(x,y)}
\newcommand{\B}[0]{B(x,y)}
\newcommand{\C}[0]{C}
\newcommand{\IAvg}[0]{\bar{I}(x,y)}
\newcommand{\nAtoms}[0]{N_\mathrm{atoms}}
\newcommand{\nImages}[0]{n_\mathrm{images}}
\newcommand{\nComps}[0]{n_\mathrm{comp}}
\newcommand{\pc}[1]{\mathrm{pc}_{#1}(x,y)}
\newcommand{\pcCorr}[1]{\mathrm{pc}'_{#1}(x,y)}
\newcommand{\coeff}[1]{w_{#1}(\idx)}
\newcommand{\coeffCorr}[1]{w'_{#1}(\idx)}
\newcommand{\idx}[0]{i}
\newcommand{\x}[0]{x}
\newcommand{\y}[0]{y}
\newcommand{\ellipseR}[0]{r}
\newcommand{\ellipseXC}[1][black]{{\color{#1}x_c}}
\newcommand{\ellipseYC}[1][black]{{\color{#1}y_c}}
\newcommand{\ellipseA}[1][black]{{\color{#1}a}}
\newcommand{\ellipseB}[1][black]{{\color{#1}b}}
\newcommand{\ellipseT}[0]{t_\idx}
\newcommand{\ellipseTheta}[1][black]{{\color{#1}\vartheta}}
\newcommand{\angleDiff}{\varphi_\mathrm{diff}}
\newcommand{\PhiOffsetDiff}[0]{\vartheta(i)}

\newcommand{\ellipseFit}[1][black]{
    \coeff{1} = \ellipseXC[#1] + \ellipseA[#1] \cos \ellipseTheta[#1] \cos \ellipseT - \ellipseB[#1] \sin \ellipseTheta[#1] \sin \ellipseT\\
    -\coeff{2} = \ellipseYC[#1] + \ellipseA[#1] \sin \ellipseTheta[#1] \cos \ellipseT + \ellipseB[#1] \cos \ellipseTheta[#1] \sin \ellipseT
}

\newcommand{\coeffCorrection}[1][black]{
    \begin{pmatrix}
        \coeffCorr{1}\\
        \coeffCorr{2}
    \end{pmatrix}
        &=
    \underbrace{
        \begin{pmatrix}
            1/\ellipseA[#1] & 0\\
            0 & 1/\ellipseB[#1]
        \end{pmatrix}
        \times
    }_\text{scale}
    \underbrace{
    \begin{pmatrix}
        \cos(-\ellipseTheta[#1]) & -\sin(-\ellipseTheta[#1])\\
        \sin(-\ellipseTheta[#1]) & \cos(-\ellipseTheta[#1])
    \end{pmatrix}
    \times
    }_\text{rotate}
    \left[
        \begin{pmatrix}
            \coeff{1}\\
            -\coeff{2}
        \end{pmatrix}
        \smash[b]{
            \underbrace{
            -
            \begin{pmatrix}
                \ellipseXC[#1]\\
                \ellipseYC[#1]
            \end{pmatrix}
            }_\text{translate}
        }
    \right]
}

\newcommand{\compCorrection}[1][black]{
    \begin{pmatrix}
        \pcCorr{1}\\
        \pcCorr{2}
    \end{pmatrix}
    &=
    \underbrace{
        \begin{pmatrix}
        \ellipseA[#1] & 0\\
        0 & \ellipseB[#1]
        \end{pmatrix}
        \times
    }_\text{scale}
    \underbrace{
        \begin{pmatrix}
        \cos(\ellipseTheta[#1]) & -\sin(\ellipseTheta[#1])\\
        \sin(\ellipseTheta[#1]) & \cos(\ellipseTheta[#1])
        \end{pmatrix}
        \times
    }_\text{rotate}
    \begin{pmatrix}
        \pc{1}\\
        \pc{2}
    \end{pmatrix}
}

\begin{document}

\title{Principal Component Analysis for Spatial Phase Reconstruction in Atom Interferometry}

\author{Stefan Seckmeyer}
    \email{seckmeyer@iqo.uni-hannover.de}
\author{Holger Ahlers}
    \altaffiliation{Present Address: DLR Institute for Satellite Geodesy and Inertial Sensing, 30167 Hannover, Germany}
\author{Jan-Niclas Kirsten-Siem\ss}
\author{Matthias Gersemann}
\author{Ernst M. Rasel}
\author{Sven Abend}
\author{Naceur Gaaloul}
    \email{gaaloul@iqo.uni-hannover.de}

\affiliation{Gottfried Wilhelm Leibniz Universit\"at Hannover, Institut f\"ur Quantenoptik, Welfengarten 1, 30167 Hannover, Germany}

\date{\today}
\begin{abstract}
Atom interferometers are sensitive to a wide range of forces by encoding their signals in interference patterns of matter waves.
To estimate the magnitude of these forces, the underlying phase shifts they imprint on the atoms must be extracted.
Up until now, extraction algorithms typically rely on a fixed model of the patterns' spatial structure, which if inaccurate can lead to systematic errors caused by, for example, wavefront aberrations of the used lasers.
In this paper we employ an algorithm based on Principal Component Analysis, which is capable of characterizing the spatial phase structure and per image phase offsets of an atom interferometer from a set of images.
The algorithm does so without any prior knowledge about the specific spatial pattern as long as this pattern is the same for all images in the set.
On simulated images with atom projection noise we show the algorithm's reconstruction performance follows distinct scaling laws, i.e., it is inversely-proportional to the square-root of the number atoms or the number of images respectively, which allows a projection of its performance for experiments.
We also successfully extract the spatial phase patterns of two experimental data sets from an atom gravimeter.
This algorithm is a first step towards a better understanding and complex spatial phase patterns, e.g., caused by inhomogeneous laser fields in atom interferometry.
\end{abstract}

\maketitle

\section{Introduction}
\begin{figure}
    \centering
    \subfloat[]{
        \centering
        \includegraphics[width=0.6\textwidth,valign=t]{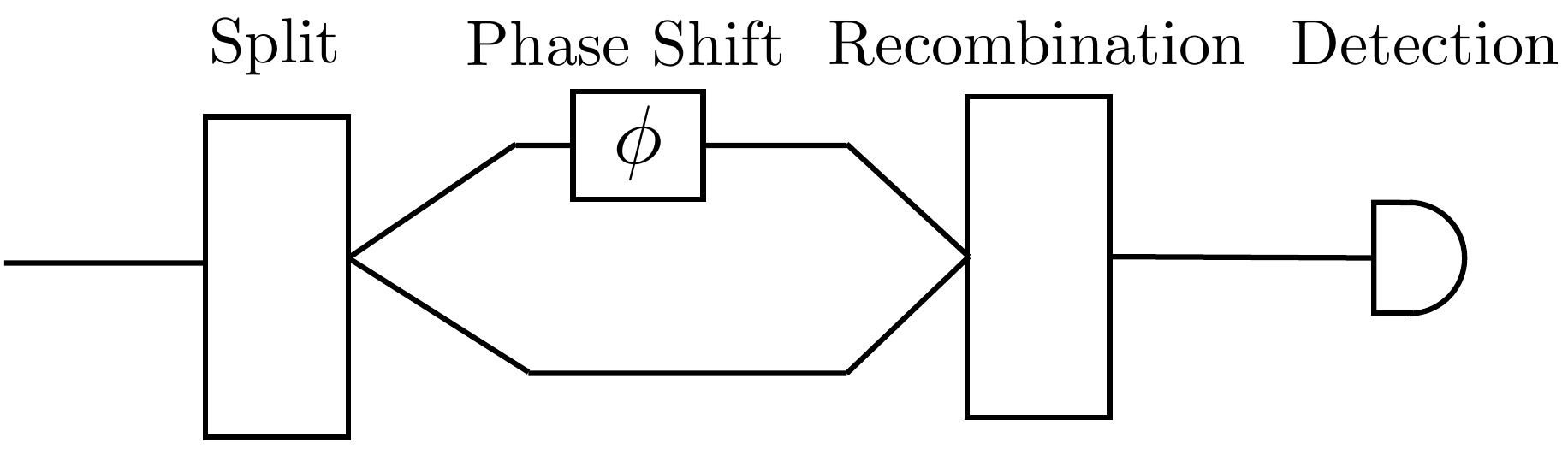}
        \label{fig:GeneralInterferometer}
    }
    \hfill
    \subfloat[]{
        \centering
        \begin{tblr}{
            colspec = {Q[c,h]Q[c,h]},
        }
            \phantom{aaaaaa}$\phi=0.4\pi$ & \phantom{aaaaaa}$\phi=2\pi$\\
            \includegraphics[width=0.3\textwidth]{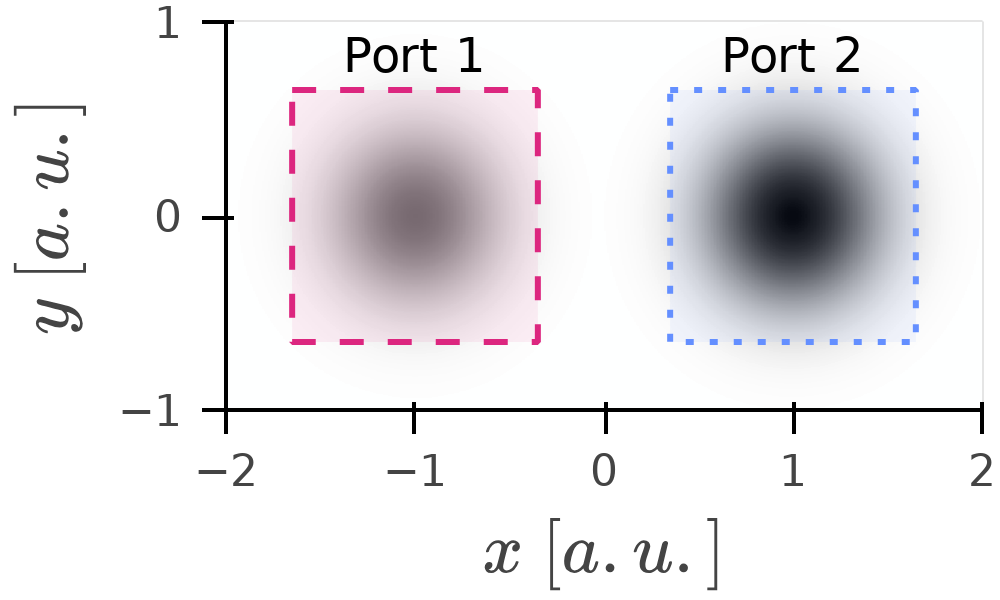}
            &\includegraphics[width=0.3\textwidth]{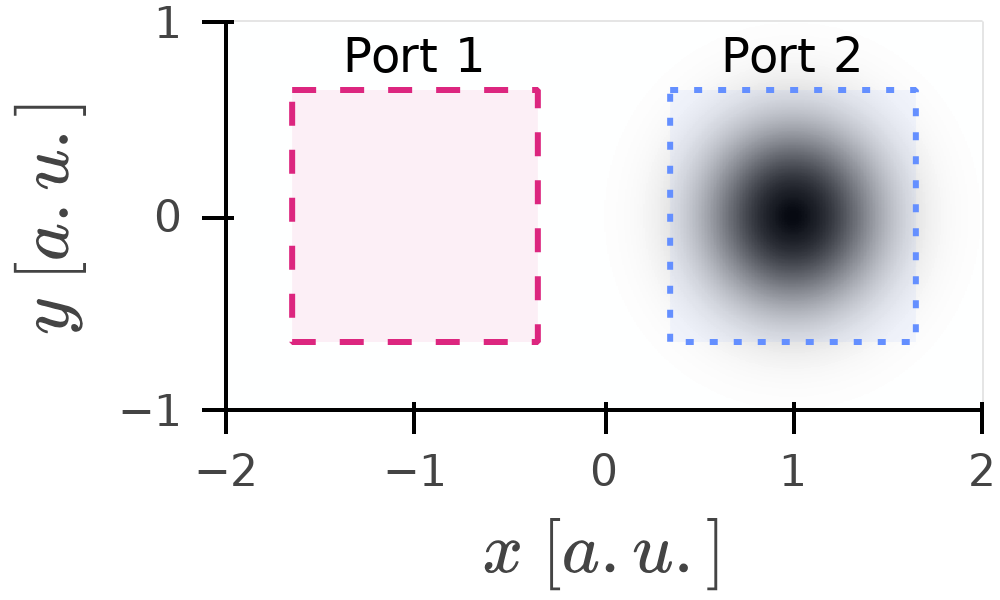}
        \end{tblr}
        \label{fig:phaseEstBoxIntegrate}
    }
    \hfill
    \subfloat[]{
        \centering
        \includegraphics[width=0.6\textwidth,valign=t]{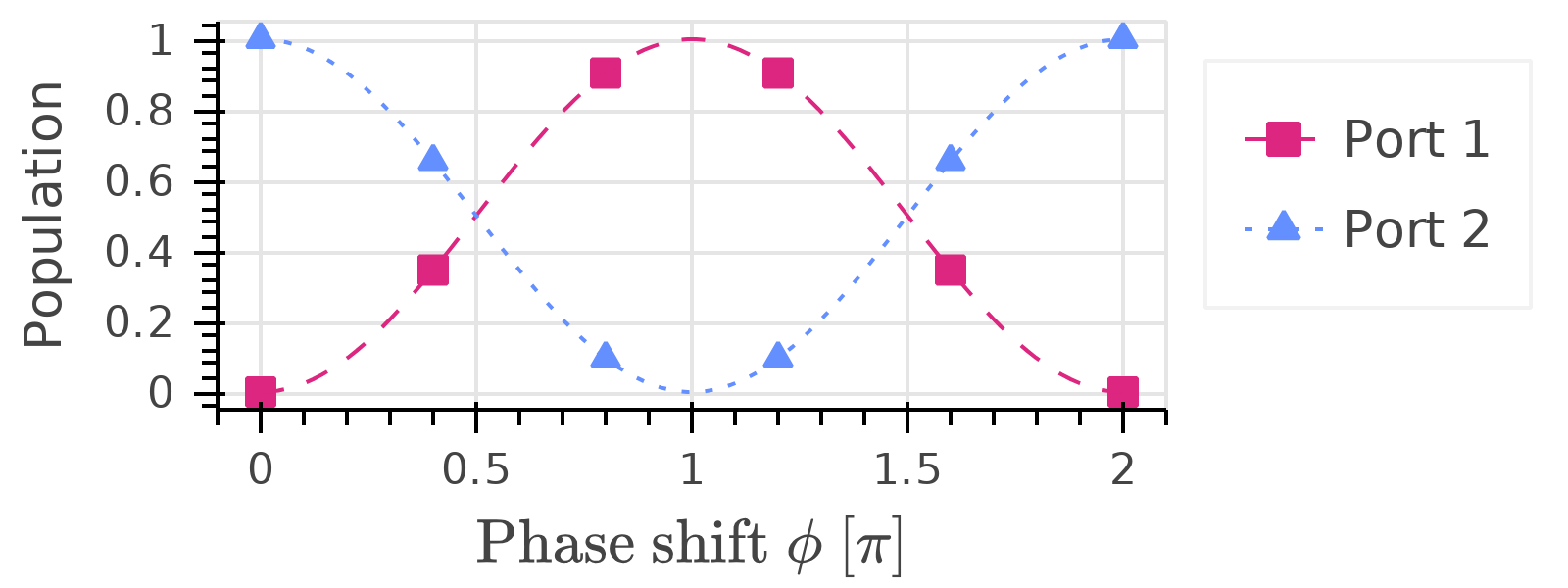}
        \label{fig:phaseEstBoxScan}
    }
    \caption{
    a) Scheme of an atom interferometer. The phase shift $\phi$ can be caused by multiple effects and is abstracted by a box on one of the arms.
    b) Ideal images of the two ports of an atom interferometer for two different phase shifts $\phi$. The ports are illustrated by the marked integration regions for simple box integrals. The integration is the estimation of the intensity in each port.
    c) A phase scan: The population of the two ports as a function of a deliberately added phase shift $\phi$. To estimate the phase shift a cosine can be fitted to this oscillation in the populations.
}
\end{figure}

Atom interferometry allows for high precision experiments in several fields such as gravimetry~\cite{Kasevich1991, Menoret2018, Wu2019}, gravity cartography~\cite{Stray2022} and inertial navigation~\cite{Geiger2011, Cheiney2018}.
Other use-cases include measurements of fundamental constants like the fine structure constant ~\cite{Morel2020, Parker2018a}, the gravitational constant~\cite{Rosi2014} and the universality of free fall~\cite{Asenbaum2020}.
In atom interferometry, a cloud of atoms is usually coherently split and recombined via atom-light interactions with laser pulses.
The signal to be measured is translated into a relative phase shift between the interferometer arms as depicted in \cref{fig:GeneralInterferometer}.
However, such a phase difference is not a quantum mechanical observable and therefore cannot be measured directly.

To measure the phase of an atom interferometer, it is projected into an interferogram, which can be detected after the recombination of the two matter waves via a final laser pulse (\cref{fig:GeneralInterferometer}).
\Cref{fig:phaseEstBoxIntegrate}  illustrates a result of such a measurement of atomic populations, e.g., obtained via fluorescence or absorption imaging \cite{Borde1989, KetterleW.1999}.
These two images also show the typical structure of such an interferogram having two so-called ports.
In a light interferometer the ports' brightness is given by the light intensity.
For atom interferometers it is determined by the number of atoms interacting with the detection laser.
Their relative brightness between the ports directly depends on the phase shift introduced during the sequence (see \cref{fig:GeneralInterferometer}).

In order to reconstruct the phase difference from its projection into intensity, one needs to recover their functional relationship.
To do so, interferograms with different phase offsets are recorded over multiple runs to cover a whole oscillation period, which results in a phase scan as shown in \cref{fig:phaseEstBoxScan}.
Alternatively, intentionally creating a phase shear imprint  across the atomic clouds and fitting a fixed model has also been used to perform a  phase measurement in a single shot \cite{Dickerson2013, Sugarbaker2013, Asenbaum2020}.

It is typically assumed that the ports of an interferometer have a phase difference of $\pi$ and that the phase in each port is spatially homogeneous.
However, there are multiple processes which can introduce an unwanted spatial phase pattern in addition to the desired signal phase into each of the ports.
Early studies on spatial phase imprints in atom interferometry \cite{Wang2005, Horikoshi2006} focused on atom-atom interactions and waveguides.
These optical and magnetic waveguides often suffer from imperfections (e.g. speckles), which can lead to unwanted spatially dependent phase patterns.
The quality of atom interferometric measurements depends directly on the quality of the light used to manipulate the atoms.
Imperfections of the optical potentials are one of the leading systematic effects of current and probably future measurements.
They are imprinted into the spatial phase profile of the final recorded interferometer images and usually referred to as wavefront aberrations \cite{Niebauer1995, Schkolnik2015, Karcher2018}.

Currently, the primary mitigation strategy is to minimize these aberrations as much as possible \cite{Geiger2020a}.
That is because they are difficult to accurately measure and post-correct for in terms of systematic shifts \cite{Karcher2018, Parker2018a, Morel2020}, which is why standard analysis algorithms do not generally take them into account. Thus, this strategy introduces significant systematic errors, if the wavefront aberrations are strong enough to have a significant impact on the measurements.

In the fields of light and acoustic interferometry a very similar problem appears in phase shifting interferometry \cite{Bruning1974, Schreiber2007, Servin2014}, which extracts the phase from a set of phase-shifted interferograms.
One way to account for the spatial phase imprints of wavefront aberrations is the use of statistical algorithms like Principal Component Analysis (PCA) and Independent Component Analysis, which have been examined in \cite{Pearson1901, Hotelling1933, Segal2010, Dubessy2014, scikit-learn}.

In the context of atom interferometry, PCA was used to enhance the contrast and shape of shear-interferometer fringes from atom interferometry output images for analysis ~\cite{Sugarbaker2013,Dickerson2013}.

In this paper, we present a PCA-based, spatial phase reconstruction (PSPR) algorithm that extracts an arbitrary spatial phase map and the phase offsets between different images of an image set.
The algorithm shares similarities with the one presented in ~\cite{Yatabe2017} and we successfully apply it to both simulated atom interferometry data as well as actual measurements.
The paper is organized as follows.
In \cref{sec:Algorithms}, we introduce the PSPR algorithm and discuss the underlying assumptions in contrast to commonly used phase extraction methods in atom interferometry.
\Cref{sec:NoNoise} establishes an accuracy baseline for PSPR by analyzing synthetic data of a two-port atom interferometer in the absence of noise.
We do so by studying synthetic interferometry images affected by three exemplary phase patterns.
Each of these patterns could be reconstructed from the synthetic images at almost the numerical machine precision.
In \cref{sec:shotNoise}, we introduce atomic shot noise into our interferometer simulation and show that the performance of the phase reconstruction scales inversely proportional with the square root of the number of atoms per image as expected for uncorrelated particles.
Finally, we showcase the application of the PSPR to experimental data in \cref{sec:expData} and show its robustness under real-world conditions.
We extract the spatial phase patterns for two different interferometer configurations and verify that the coefficients exhibit structures similar to the results from the simulations.
To conclude, in \cref{sec:Discussion} we discuss the potential of PSPR for precision interferometry in general and for automation and data compression in particular for transportable atom interferometry sensors deployed on ground or in space.

\section{PCA-based Spatial Phase Reconstruction (PSPR) Algorithm}\label{sec:Algorithms}

\subsection{The Interferometer Models Underlying the Extraction Algorithms}
Every phase extraction algorithm is based on an assumed model of how the interferometer works.
Any deviation of the actual experiment from that model causes a systematic error in the results of the algorithm.

In almost all phase extraction algorithms, one needs to vary the phase difference of an interferometer to reconstruct the phase that was projected into the interferograms.
Varying the phase difference between the two arms causes the intensity to oscillate, which allows to distinguish between background intensity and the intensity of the interferometer output ports.
This phase is usually varied between different experimental runs and therefore images.
It is also possible to vary the phase via an imprint across the atomic cloud to allow the phase extraction from a single image, as done in phase shear interferometry \cite{Sugarbaker2013, Dickerson2013}.

As a baseline for the models typically used in current state-of-the-art atom interferometers \cite{Berman1997, HoganJ.M.2009}, we focus on taking multiple images and changing a phase offset in the last beam splitter.
Assuming the phase in each of the ports is constant over the whole area of the port, the phase offset determines how many atoms end up in the left or right port as shown in \Cref{fig:phaseEstBoxIntegrate}.
In order to extract amplitudes from these images, the ratio of the number of atoms in each port needs to be determined.
One approach is to simply define an area in the image and integrate the intensity inside it, to determine the ratio of atom numbers between the areas (see \cref{fig:phaseEstBoxIntegrate}).
Defining this integration area is, however, a difficult trade-off because the clouds become very dilute at the edges.
On the one hand, choosing a too large integration area reduces the signal-to-noise ratio because the outer areas of the clouds are dominated by background noise.
On the other hand, if the integrated region is too small, information is discarded.
Selecting a shape for the integration area that more closely reflects the spatial distribution of the atoms can provide a better alternative that allows using the whole atom cloud.
This, however, can introduce an additional systematic error, if the assumed cloud shape deviates from the actual data.

Both the box integral as well as fitting a specific shape on each cloud rely on estimating the number of atoms in each port, also called populations.
Plotting these populations as a function of the phase shift in the last beam splitter results in oscillations like the ones depicted in \cref{fig:phaseEstBoxScan}.
The populations of each port are fitted to a cosine, in order to determine the contrast and the phase shift of the interferometer.
This population oscillation is shifted by the phase shift caused by the measured quantity.
Each population value can only be uniquely mapped into an interval from $[0,\pi]$ which results in a dynamic range of $\pi$.
Doing such a cosine fit requires to reliably determine and control the phase shift of each image relative to the other images or to fix the ambiguity introduced by phase jumps by correlating the atom interferometer with a classical interferometer \cite{Geiger2011, Merlet2009, Geiger2020a}.

PSPR, on the other hand, does not rely on the existence of a port structure at all.
It only requires a minimum of two pixels which have a fixed phase offset $\neq \pi$ between each other over the whole image sequence.
This is formalized as defining a set of images that are assumed to follow the form
\begin{align}
    \I &= \A +\B \ \cos[\phiIXY)] \label{eq:imgModel}\\
    \phiIXY &= \phiI + \phiXY.
\end{align}
The image set has $\nImages$ images which are indexed with $\idx$ and each image has $\x \y$ many pixels with $x$ indexing the pixel row and $y$ indexing the pixel column.
$\A$ is the background intensity, $\B$ is an intensity envelope defining the signal strength of each pixel and $\phiXY$ is the free-form spatial phase profile of the interferometer.
The intensities and the phase profile are assumed to be the same in all analyzed images in the input set.
$\phiI$ is the sum of the phase shift from the measured quantity and the artificially introduced phase shift in image $\idx$.

The form of \cref{eq:imgModel} assumes that the intensity of each pixel is only ever influenced by a single phase $\phiIXY$, so it requires non-overlapping ports.
Both the intensity and the phase do not, however, have any predetermined spatial form and the order in which the images are recorded does not matter either.
From such a set of images, PSPR allows to reconstruct $\phiI$ and $\phiXY$ with a dynamic range of $2\pi$.

\subsection{PCA Based Spatial Phase Reconstruction (PSPR)}

\begin{figure}
    \centering
    \includegraphics[width=13.5cm]{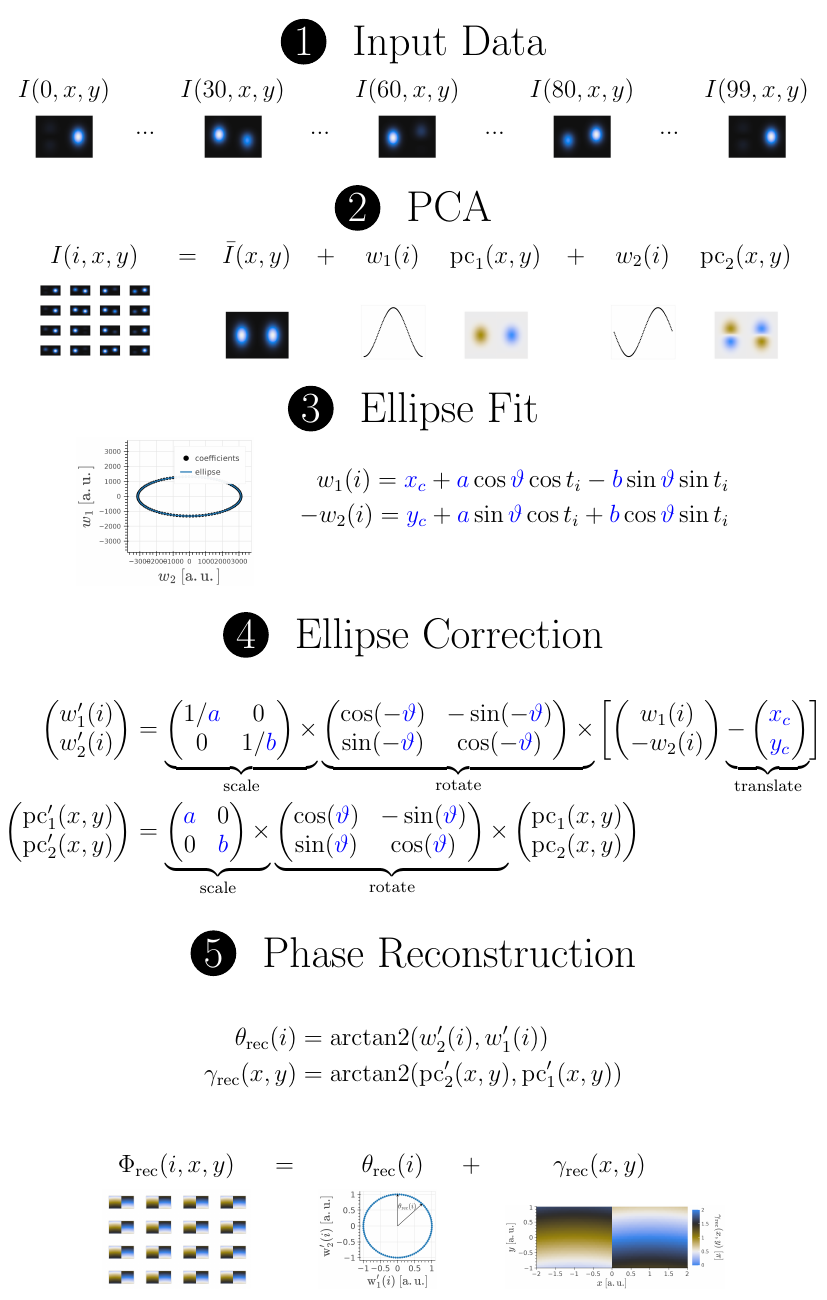}
    \caption{Steps of the PSPR algorithm.
    Analysis begins with a set of images $\I$ (1), those are transformed with PCA (2), an ellipse is fit to the coefficients $\coeff{1}$ and $\coeff{2}$ (3), the coefficients $\coeff{j}$ and components $\pc{j}$ are corrected for the ellipse distortion (4) and the phase is extracted via arctan2 (5).}
    \label{fig:ComputeGraph}
\end{figure}

The PSPR reconstruction steps are visually displayed in \cref{fig:ComputeGraph} and follow a similar logic than the one of reference \cite{Yatabe2017}.
In the following, we introduce subscripts to properly distinguish the input phases $\phiIInp$ and $\phiXYInp$ from the reconstructed phases $\phiIRec$ and $\phiXYRec$.
The input model of \cref{eq:imgModel} can be rewritten as a Fourier decomposition yielding linear combinations of terms depending only on $i$ or $x,y$:
\begin{align}
    \I = \A &+ \B \cos[\phiIInp] \cos[\phiXYInp]- \B \sin[\phiIInp] \sin[\phiXYInp]. \label{eq:linearDecomp}
\end{align}

As shown above, the image model has two linearly independent degrees of freedom.
These two degrees of freedom can be extracted using PCA.
It decomposes any data set into the average intensity $\IAvg=\mathrm{mean}[\I]$ and $\nComps$ linearly independent orthonormal components $\pc{j}$ with associated coefficients $\coeff{j}$.
As depicted in step 2 in \cref{fig:ComputeGraph}, PCA generates only two $\pc{j}$ and $\coeff{j}$ respectively, since there are only two linearly independent components in any set of images generated with \cref{eq:imgModel}:
\begin{align}
    \I = \IAvg + \coeff{1} \pc{1} + \coeff{2} \pc{2}.
\end{align}
The principal components $\pc{j}$ are chosen by PCA such that they are orthonormal and maximize the variance of the coefficients $\coeff{j}$.
Although $\cos[\phiXYInp]$ and $\sin[\phiXYInp]$ are orthogonal to each other analytically, they are not orthogonal for almost any choice of $\A$, $\B$, $\phiIInp$ and $\phiXYInp$.
Likewise, the mean $\IAvg$ does not generally equal $\A$.

This causes plotting $\coeff{1}$ against $\coeff{2}$ or $\pc{1}$ against $\pc{2}$ (with $\B=1$) to result in ellipses instead of the ideal circles.
If $\B \neq 1$ the ellipse formed by $\pc{1}$ and $\pc{2}$ becomes distorted because each point has a potentially different distance from the origin.
Therefore, an ellipse fit \cite{Halir1998a, scikit-image} is performed on the generated $\coeff{1}$ and $\coeff{2}$ instead of fitting $\pc{1}$ and $\pc{2}$ as done in \cite{Yatabe2017}.

The ellipse fit (step 3 in \cref{fig:ComputeGraph}) has the free fit parameters $\ellipseXC$, $\ellipseYC$, $\ellipseA$, $\ellipseB$ and $\ellipseTheta$:
\begin{align}
    \ellipseFit
\end{align}
The negation in front of $\coeff{2}$ stems from the subtraction in \cref{eq:linearDecomp}.
This negation needs to be assigned to either the coefficients or the components during reconstruction in order to have a consistent sign between $\phiIRec$ and $\phiXYRec$.

Correcting for the ellipse distortion in step 4 transforms $\coeff{1}$ and $\coeff{2}$ onto an approximate unit circle.
\begin{align}
    \coeffCorrection\\
    \compCorrection
\end{align}

Computing the angle of these corrected components $\pcCorr{1}$ and $\pcCorr{2}$ and coefficients $\coeffCorr{1}$ and $\coeffCorr{2}$ on their respective unit circle in step 5 reconstructs the full phase $\phiIXYRec$ for each image:
\begin{align}
    \phiIRec &= \mathrm{arctan2}(\coeffCorr{2}, \coeffCorr{1})\\
    \phiXYRec &= \mathrm{arctan2}(\pcCorr{2}, \pcCorr{1})\\
    \phiIXYRec &= \phiIRec + \phiXYRec.
\end{align}

In addition to the already mentioned \cite{scikit-image,scikit-learn} the implementation of the algorithm, simulation and visualization in this paper use the python libraries numpy \cite{harris2020array} and jax \cite{Frostig2018, jax2018github}. The bokeh library \cite{bokeh2013github} was used for plotting, colorcet \cite{colorcet2016github} and palettes from \cite{nichols2024} were used as a basis for the color schemes.

\subsection{Comparison between Inputs, Outputs and Multiple Data Sets}
The first fundamental building block for phase comparisons is the difference between two angles.
When viewing the two angles on a circle there are two distances between them, one going left-handed and the other going right-handed.
Given two phases $\varphi_1$ and $\varphi_2$, \cref{eq:angleDiff} returns $\angleDiff \in [-\pi, \pi)$ which is defined as the shorter of the two possible distances and the sign indicating which direction was taken around the circle:
\begin{align}
    \angleDiff(\varphi_1, \varphi_2) = \mathrm{mod}(\varphi_1 - \varphi_2, 2\pi)-\pi \label{eq:angleDiff}
\end{align}
with $\mathrm{mod}$ indicating the modulo operation.

Additionally, the model underlying PSPR (see \cref{eq:mainModel}) maps multiple phases to the same images.
This needs to be taken into account when comparing inputs and outputs or outputs from two different analysis runs.
When comparing $\phiIXYInp$ and $\phiIXYRec$, there may be a sign switch which still results in the same images:
\begin{align*}
    \cos[\phiIXYInp] &= \cos[s \ \phiIXYInp] = \cos[\phiIXYInpD] \ \ \forall s \in \{-1, 1\}.
\end{align*}

Comparing the individual outputs $\phiIRec$ and $\phiXYRec$ with the inputs $\phiIInp$ and $\phiXYInp$ adds another degree of freedom $c$ while all phases $\phiIInpD$ and $\phiXYInpD$ remain a solution for the same image set:
\begin{align}
    \begin{split}
    \cos[\phiIXYInp] &= \cos[\phiIInp + \phiXYInp]\\
    &= \cos(s\{[\phiIInp - c] + [\phiXYInp + c]\})\\
    &= \cos[\phiIInpD + \phiXYInpD]\\
    \forall \ \ c &\in [-\pi,\pi), \ s \in \{-1, 1\}.
    \end{split}
\end{align}

The constants $c$ and $s$ are unique for each run and determined by the PCA.
Both constants are determined by the signs of the eigenvalues when computing the components and the maximation of variance under orthogonality.

In the following, $c$ and $s$ will be computed from $\phiIInp$ and $\phiIRec$ as the solution to:
\begin{align}
    \min_{c,s} &\left( \sum_i\left\{ \angleDiff\left[\phiIRec, s \, \phiIInp - c\right]\right\}\right) \textrm{with} \ \ c \in [-\pi,\pi), \ s \in \{-1, 1\}.
\end{align}
This minimization problem can be solved directly without iteration as described in the appendix.

Since $c$ and $s$ are the same for both $\phiIRec$ and $\phiXYRec$, they are used for the computation of both $\phiIDiff$ and $\phiXYDiff$:
\begin{align}
    \phiIDiff &\coloneqq \angleDiff\left[s \, \phiIInp - c, \phiIRec \right]\\
    \phiXYDiff &\coloneqq \angleDiff\left[s \, \phiXYInp + c, \phiXYRec \right].
\end{align}

\FloatBarrier

\section{Noise Free Phase Reconstruction}
\label{sec:NoNoise}
This section establishes an accuracy baseline for the performance of PSPR by analyzing computer-generated sample images, inspired by a simplified model of a two-port atom interferometer.
The two ports of the interferometer have a Gaussian intensity profile $g(x,y,\sigma)$ and the same spatial phase pattern $\phiXYInp$.
They are shifted relative to each other in position by $2 \portOffset$ in x-direction, do not exhibit any overlap and the left port has an additional phase shift of $\pi$. This is parametrized by:
\begin{align}
    g(x,y,\Delta) &= \frac{1}{2 \pi \Delta^2} \mathrm{exp}\left(\frac{-(x^2 + y^2)}{2 \Delta^2}\right)\\
    \phiIXYInp &= \phiIInp + \phiXYInp \label{eq:inpSum}\\
    \I &=
    \begin{cases}
        \frac{1}{2} \, \text{d}x \, \text{d}y  \,  g(x - \portOffset, y, 0.3) \{1+ \C \cos[\phiIXYRInp]\} & \text{if } x \geq 0\\
        \frac{1}{2} \,  \text{d}x \, \text{d}y \,  g(x + \portOffset, y, 0.3) \{1+ \C \cos[\phiIXYLInp + \pi]\} & \text{if } x < 0\\
    \end{cases}.
    \label{eq:mainModel}
\end{align}
$\C$ is the contrast and $\text{d}x$ and $\text{d}y$ are the width and height of each pixel respectively.
This image model was used to generate 100 images with $\nImages=100$ each for three choices for $\phiXYInp$ with the parameters:
\begin{align}
    \phiIInp&=\frac{ 2\pi i}{\nImages} \label{inputPhiNoiseFree} \ \text{with} \ i \in [0,\nImages)\\
    \phiXYInp&= \begin{cases}
        2 x^2+4y\\
        4x+4y\\
        4 (0.2x)^2+2y
    \end{cases}
    .\label{eq:inputVarPhiNoiseFree}
\end{align}
The input phase patterns $\phiXYInp$ and the main result of this section $\phiXYDiff$ are plotted in \cref{fig:NoNoisePhiXY}.
$\phiXYDiff$ is, with the exception of a few outlier pixels, close to machine precision.
The differences vary slightly from run to run and one can see weak remnants of the patterns $\phiXYInp$ in $\phiXYDiff$.
The PSPR reconstruction yields an excellent agreement in this case where $\phiIDiff$ is close to the precision of the used floating point numbers, so for double precision about $\qty{e-16}{\radian}$.

\newcommand{\phiXYTableHeightA}{2.6cm}
\newcommand{\phiXYTableHeightB}{2.88cm}
\newcommand{\phiXYTableRowTitle}[2]{ {\Large #1) $\phiXYInp = #2$} }
\begin{figure}
    \centering
    \begin{tblr}{
        colspec = {Q[c]Q[c,h]Q[c,h]},
    }
        \Large a) & \large $ 2 x^2+4y$ &\\
        &\includegraphics[height=\phiXYTableHeightA]{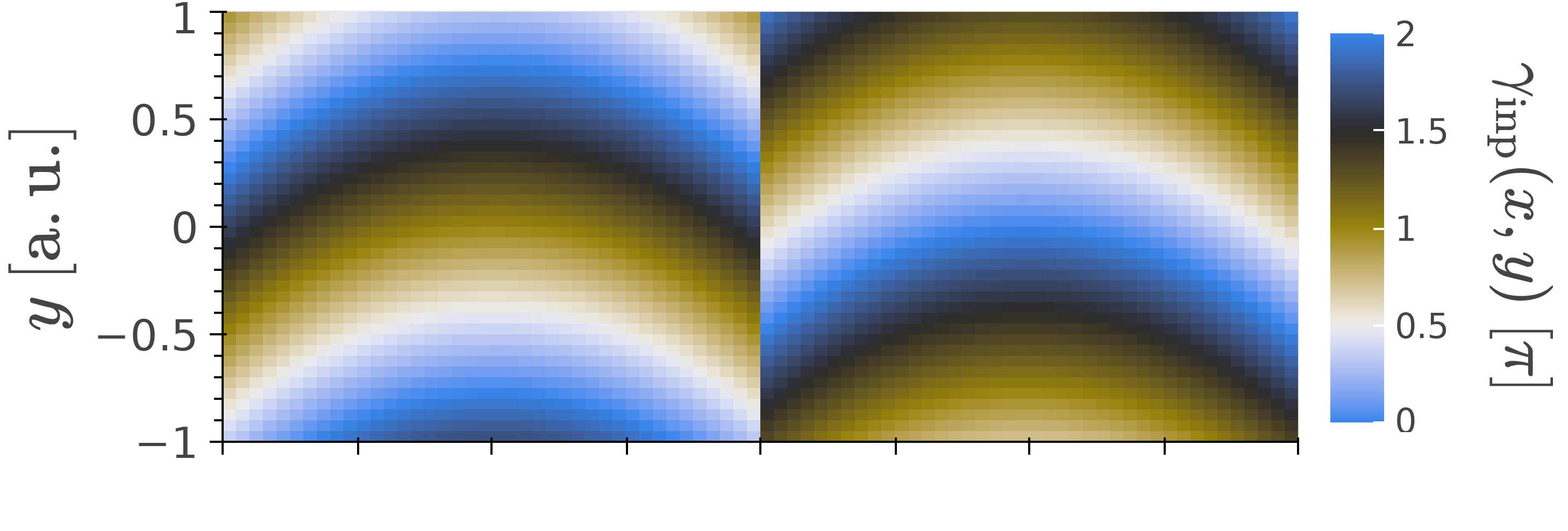}
        &\includegraphics[height=\phiXYTableHeightA]{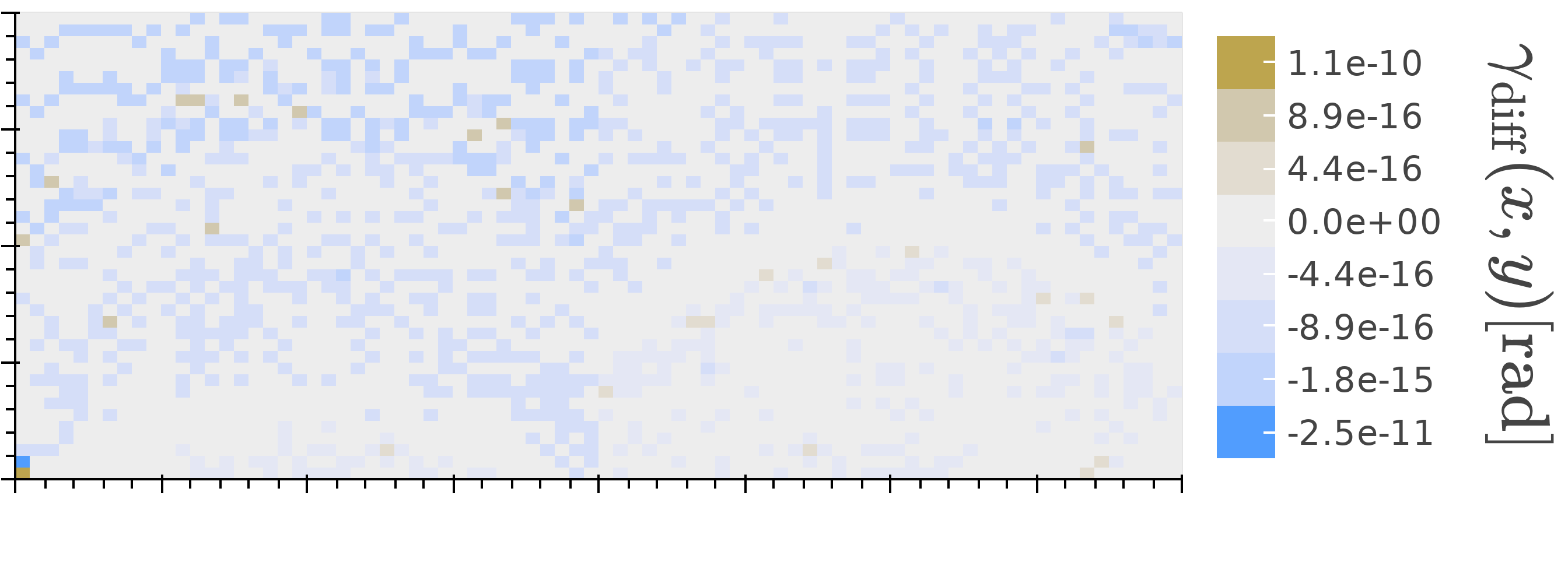}
        \\
        \Large b) & \large $4x+4y$ &\\
        &\includegraphics[height=\phiXYTableHeightA]{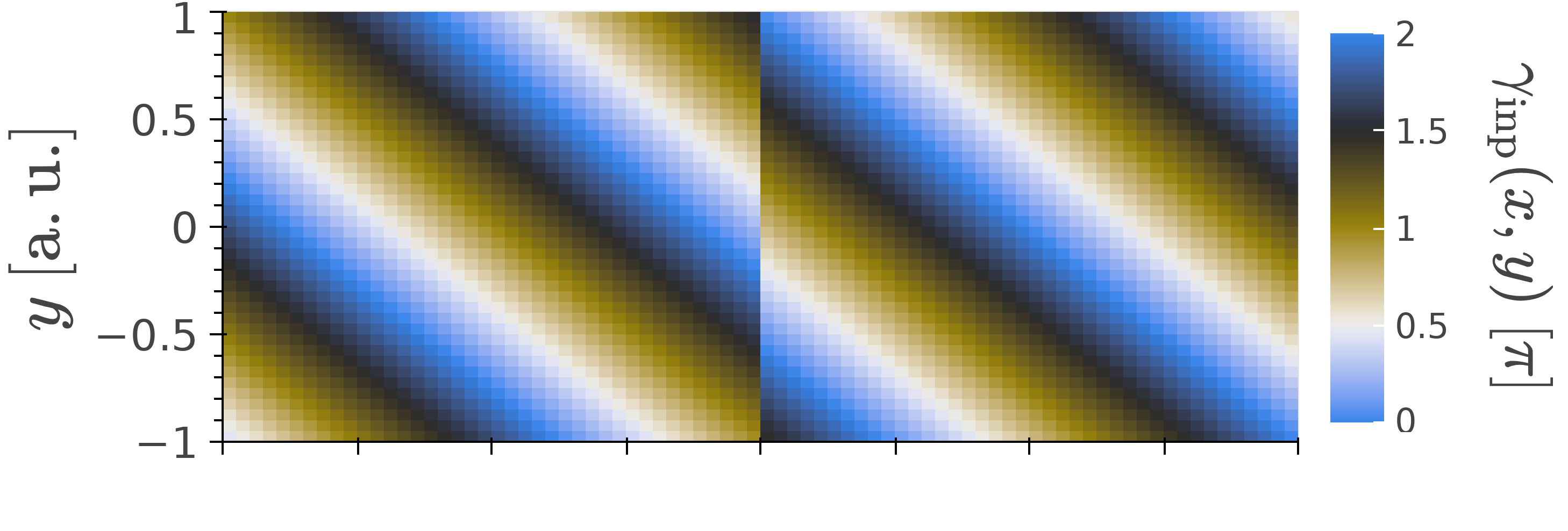}
        &\includegraphics[height=\phiXYTableHeightA]{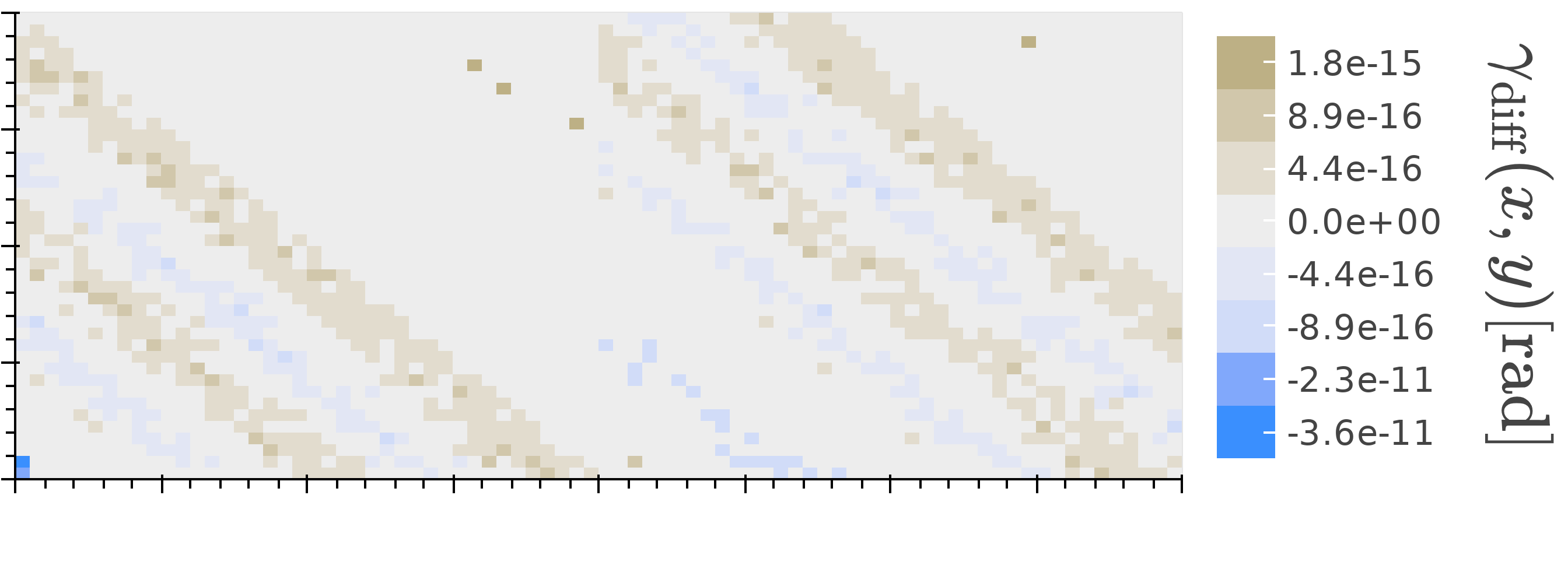}
        \\
        \Large c) & \large $4 (0.2x)^2+2y$ &\\
        &\includegraphics[height=\phiXYTableHeightB]{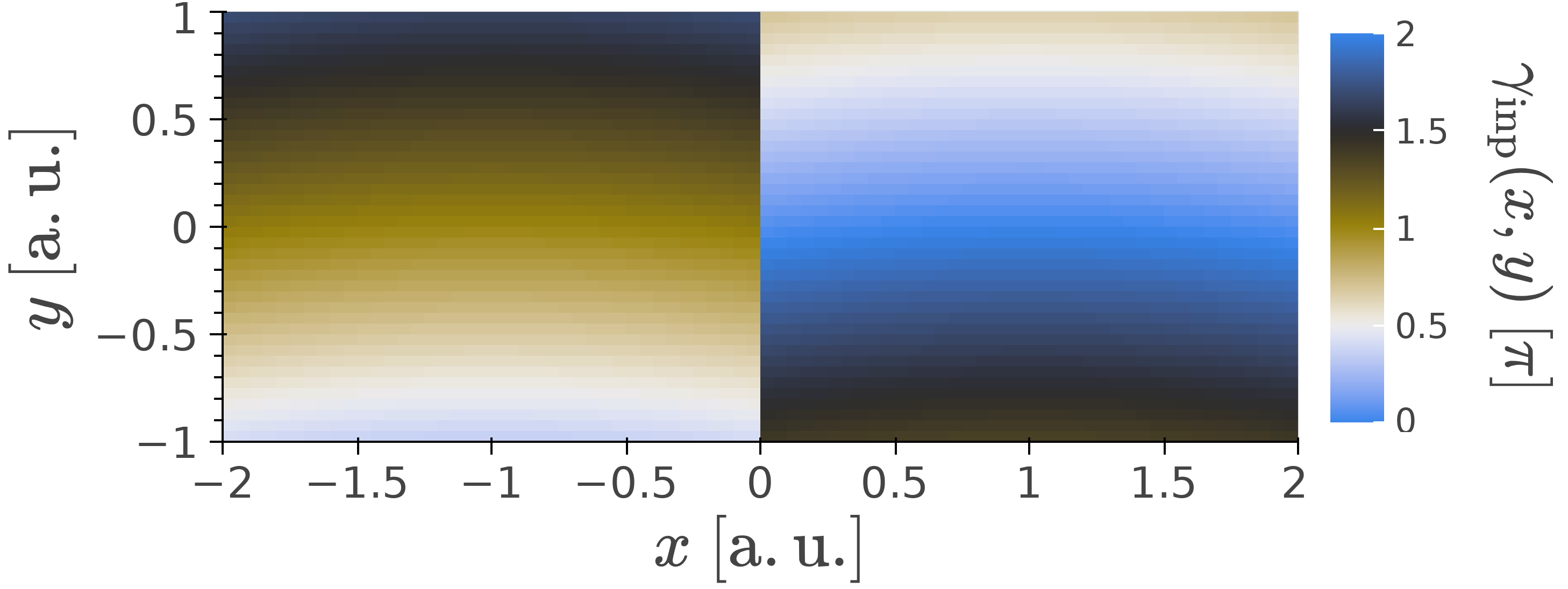}
        &\includegraphics[height=\phiXYTableHeightB]{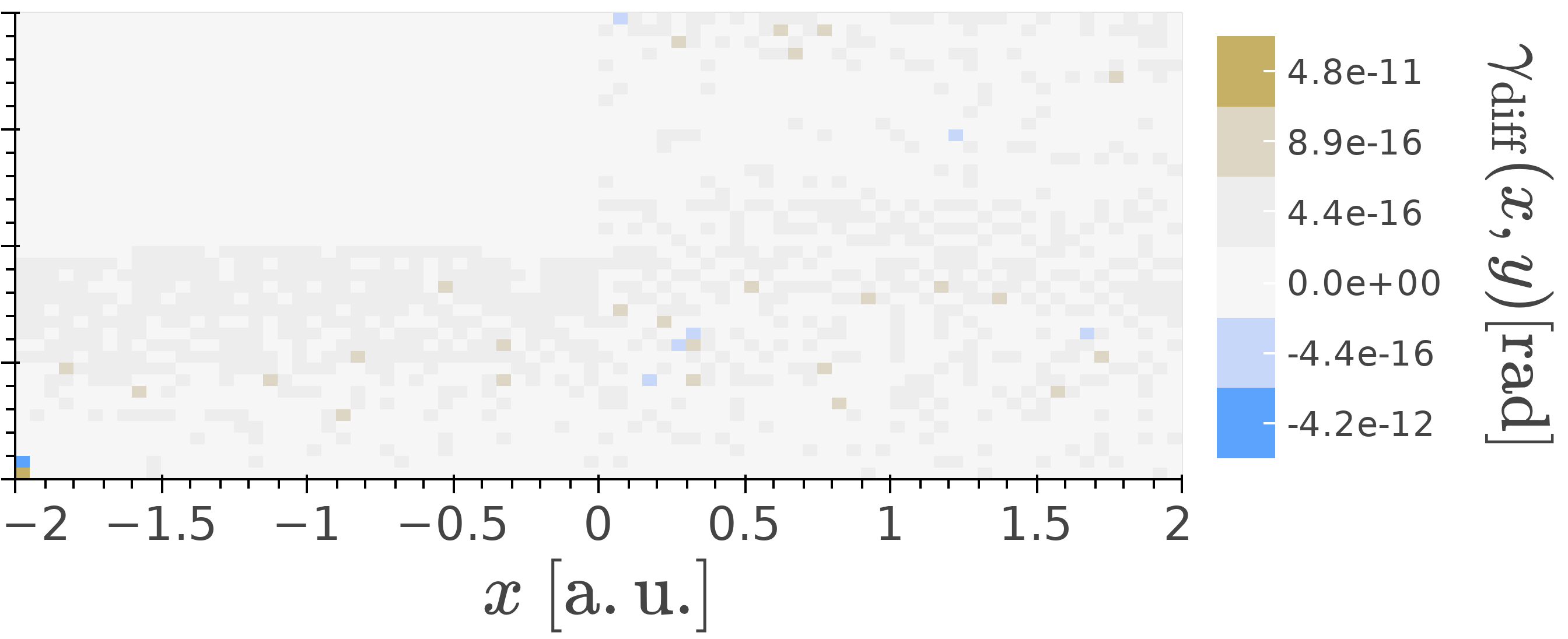}
    \end{tblr}
    \caption{
        Spatial phase patterns $\phiXYInp$ used for the synthetic image generation as defined in \cref{eq:inputVarPhiNoiseFree}. These cases were chosen to show different scaling behaviors when including noise in \cref{sec:shotNoise}. The two columns in each image correspond to the two ports of the interferometer which are modeled as the two cases in \cref{eq:mainModel}.
        The reconstruction was done from a data set of 100 images generated with \cref{eq:mainModel} and the parameters in \cref{inputPhiNoiseFree} and \cref{eq:inputVarPhiNoiseFree}.
        Almost all pixels are reconstructed close to machine precision around $\qty{e-16}{\radian}$, the original phase patterns are weakly visible.
        Note the discrete and non-linear color bars in the difference plots to accommodate outliers.
    }
    \label{fig:NoNoisePhiXY}
\end{figure}

\FloatBarrier

\section{Phase Reconstruction in the Presence of Shot Noise}
\label{sec:shotNoise}
Shot noise, also referred to as quantum projection noise~\cite{Itano1993}, is a type of noise which is present in all quantum measurements.
This type of noise was chosen to evaluate the reconstruction accuracy of PSPR in the case of realistic measurement in a quantum system.

The intensity of the model images in the previous section is used as the probability density to find an atom in a particular pixel.
For each atom, a single pixel is picked randomly according to this density.
The brightness of each pixel is then defined as the number of atoms that populate it.
All intensity values in the following graphs are therefore in units of atoms.

To understand the reconstruction performance of PSPR in the presence of shot noise, the image generation process is repeated $n_\mathrm{run}$ times.
All graphs in this chapter focus on the standard deviation $\mathrm{std}$, because the differences between input and reconstruction average to zero for the chosen parameter regimes.
Each of the following figures was generated with $n_\mathrm{run}=74000$ simulation runs per input parameter combination.
The default input parameters are $\phi(i)=2\pi i/1000$ with $i \in [0, 1000)$, $\nAtoms = 10^5$, $\nImages=1000$ and the contrast $\C=1$.

\begin{figure}
    \centering
    \subfloat[]{
        \includegraphics[height=3.2cm]{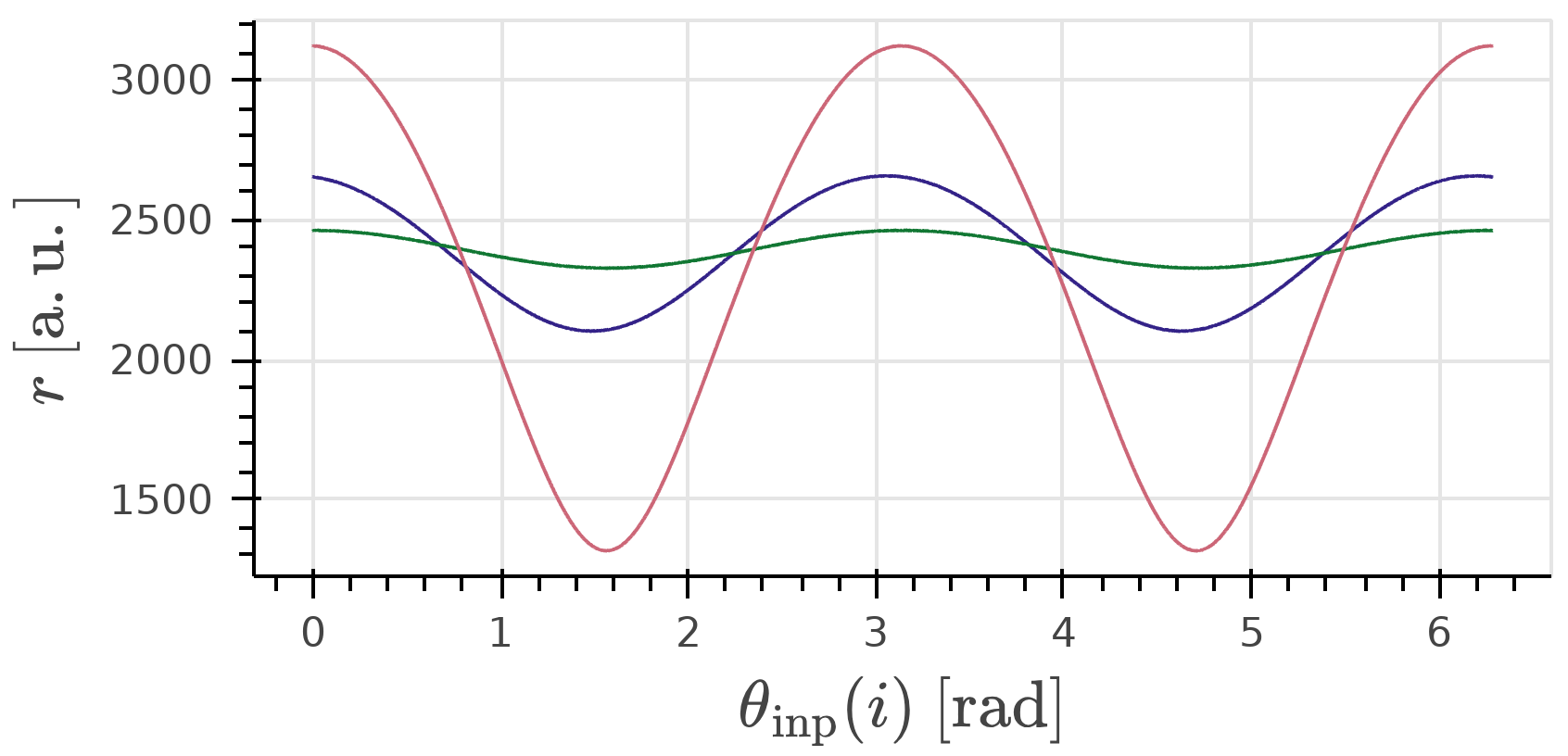}
        \label{fig:ROverVarphiI}
    }
    \subfloat[]{\includegraphics[height=3.2cm]{
        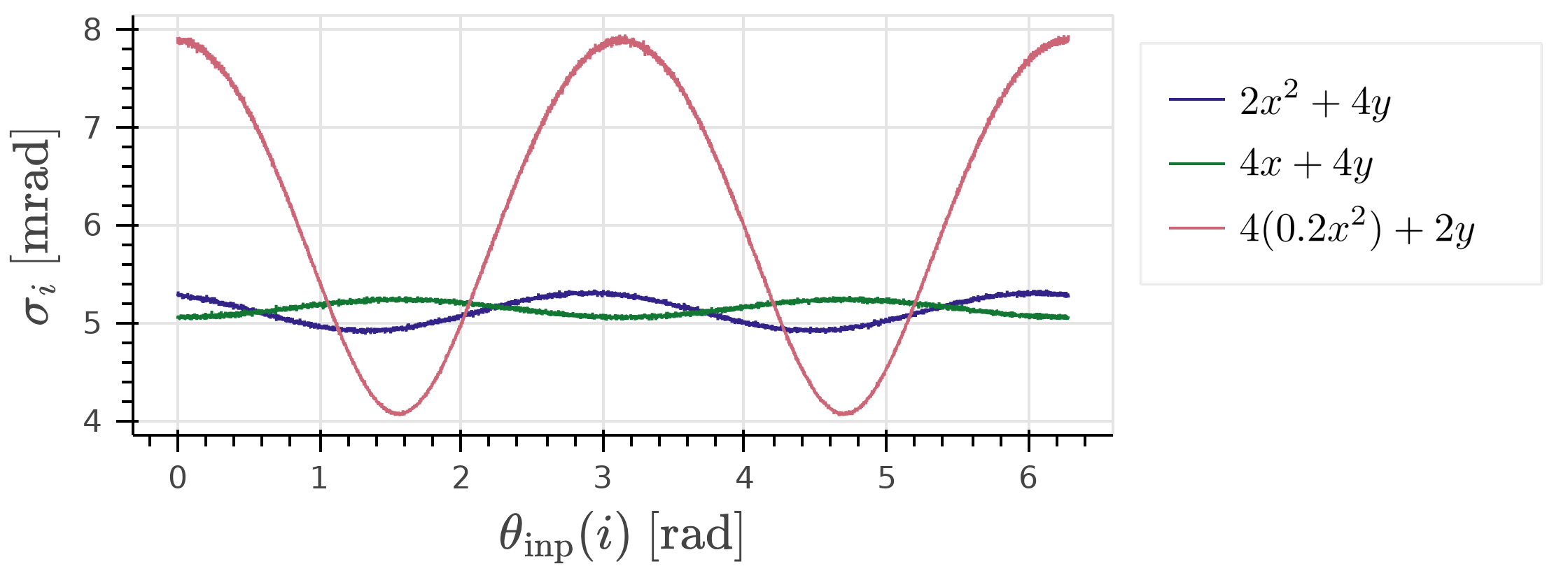}
        \label{fig:PhiIScalingNImagesPerImage}
    }
    \caption{
        Results for interferograms defined in \cref{eq:mainModel} with shot noise for $\nImages=10000$, $\nAtoms=10^5$, $\C=1$.
        a) The distance $\ellipseR$ of the coordinate $(\coeff{1}, \coeff{2}$ to the center of the fitted ellipse as a function of $\phiIInp$.
        b) The reconstruction accuracy $\phiIStd$ as a function of $\phiIInp$.
        One can clearly see the strong correlation between $\phiIStd$ and $\ellipseR$.
    }
    \label{fig:PerImageStdCombi}
\end{figure}

The reconstruction accuracy $\phiIStd \coloneqq \mathrm{std}[\phiIDiff]$ varies as a function of $\phiIInp$ (\cref{fig:PhiIScalingNImagesPerImage}).
This strongly correlates with the radius of the coefficient pair on the ellipse (\cref{fig:ROverVarphiI}).

This is abstracted away by taking the maximum $\phiIStdMax$, in order to understand and characterize the shot-noise scaling as a function of other simulation parameters.
$\phiXYStd \coloneqq \mathrm{std}[\phiXYDiff]$ does not show any scaling as a function of $\phiIXYInp$.

When systematically changing the number of atoms per image $\nAtoms$, the number of images $\nImages$ per analysis, the contrast $C$, and the spatial phase distribution $\phiXYInp$ one finds the following empirical scaling laws for $\phiIStdMax$ and $\phiXYStd$:
\begin{align}
    \phiIStdMax &= \frac{a}{\C \ \nAtoms^{1/2}} \label{eq:shotNoiseI}\\
    \phiXYStd &= \frac{b}{C \ \left[\IAvg \ \nImages\right]^{1/2}}  \label{eq:shotNoiseXY}.
\end{align}
$a$ and $b$ are fit constants with $a$ depending on the chosen pattern $\phiXYInp$:
\begin{align*}
    a &\approx
    \begin{cases}
        1.70 \ &\phiXYInp = 2x^2+4y\\
        1.67 \ &\phiXYInp = 4x+4y\\
        2.51 \ &\phiXYInp = 4(0.2 x^2)+2y
    \end{cases}\\
    b &\approx 1.41.
\end{align*}

Both laws match the intuition that the reconstruction accuracy depends on how many interfering atoms were observed per reconstructed value and are compatible with atomic shot noise scaling.

\begin{figure}
    \centering
    \begin{minipage}[t]{0.48\textwidth}
        \includegraphics[width=\textwidth]{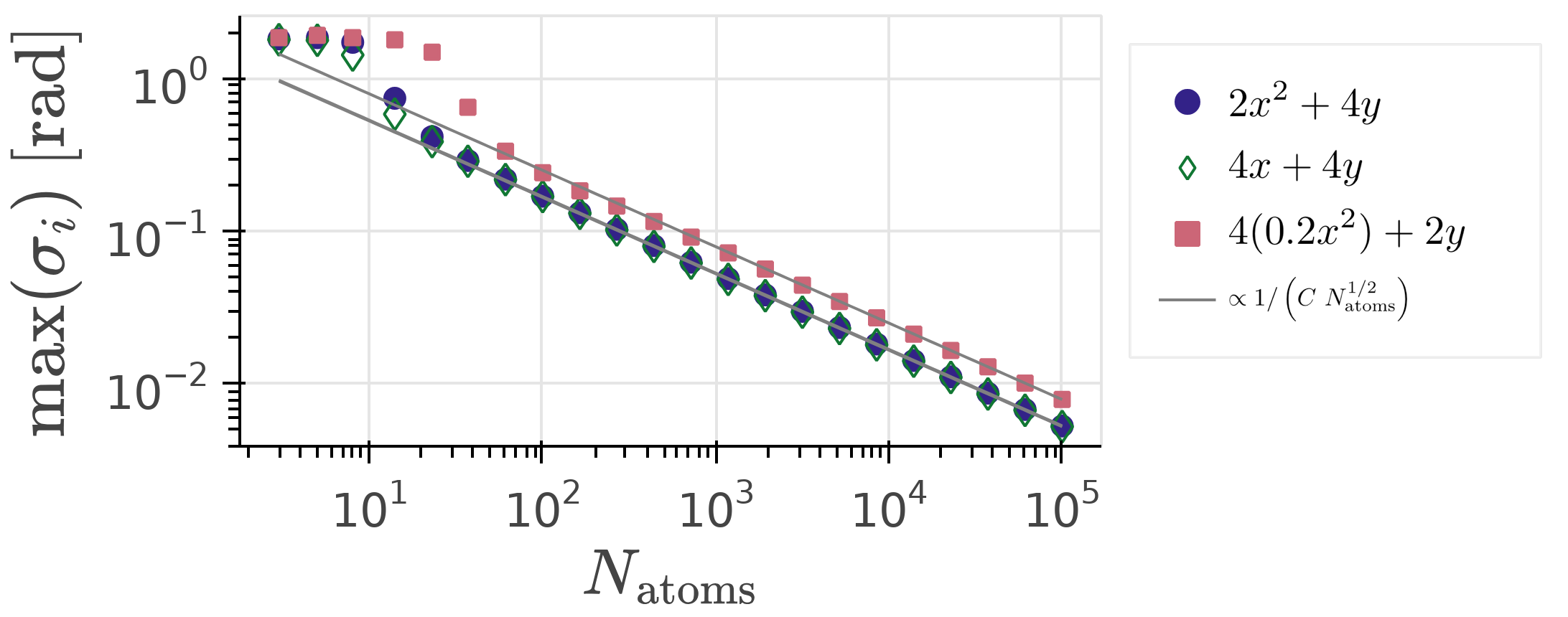}
        \caption{
            Worst-case reconstruction accuracy $\phiIStdMax$ as a function of $\nAtoms$ for $C=1$, $\nImages=1000$ and $\phiIInp$ as in \cref{inputPhiNoiseFree}.
            For approximately $\nAtoms>100$ PCA works well enough so that $\phiIStdMax$ scales with $a/(C \cdot \sqrt{N_\mathrm{atoms}})$, with $a$ being a fit constant which depends on spatial phase pattern $\phiXYInp$.
        }
        \label{fig:PhiIScalingNAtoms}
    \end{minipage}
    \begin{minipage}[t]{0.48\textwidth}
        \includegraphics[width=\textwidth]{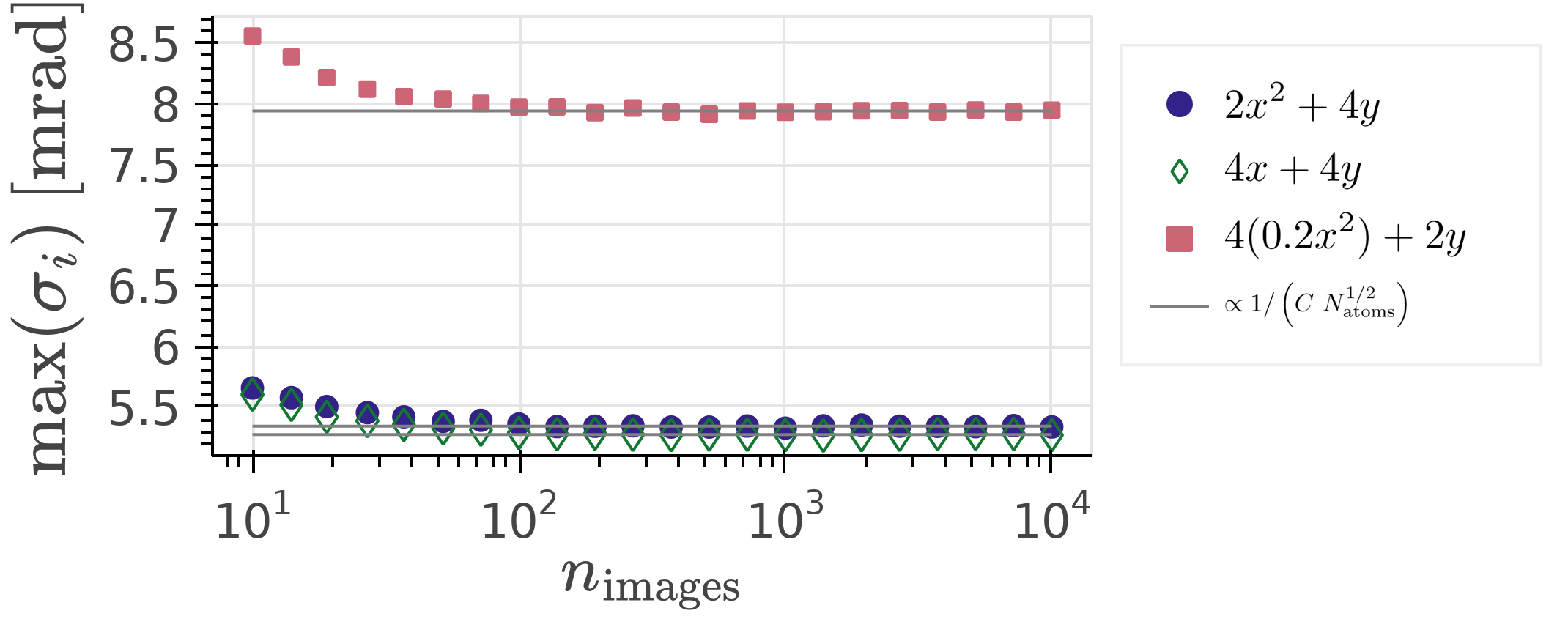}
        \caption{
            Worst-case reconstruction accuracy $\phiIStdMax$ as a function of $\nImages$.
            $\phiIStdMax$ is independent of $\nImages$ and therefore only improves until $\nImages=50$ which are enough images for PSPR to work well for $C=1$, $\nAtoms=10^5$ and $\phiIInp$ as in \cref{inputPhiNoiseFree}.
        }
        \label{fig:PhiIScalingNImages}
    \end{minipage}
\end{figure}

\Cref{fig:PhiIScalingNAtoms} and \cref{fig:PhiIScalingNImages} confirm the shot-noise scaling of \cref{eq:shotNoiseI} and \cref{eq:shotNoiseXY} with $\nAtoms$ and $\nImages$, respectively.
One can see that for about $\nAtoms>100$ atoms per image and for $\nImages>50$ images per sequence the signal quality is high enough for the scaling law \cref{eq:shotNoiseI} to hold.
Since $\phiIStdMax$ does not scale as a function of $\nImages$ the reconstruction accuracy converges towards constant values for large enough image sets.

\begin{figure}
    \centering
    \includegraphics[width=0.58\textwidth]{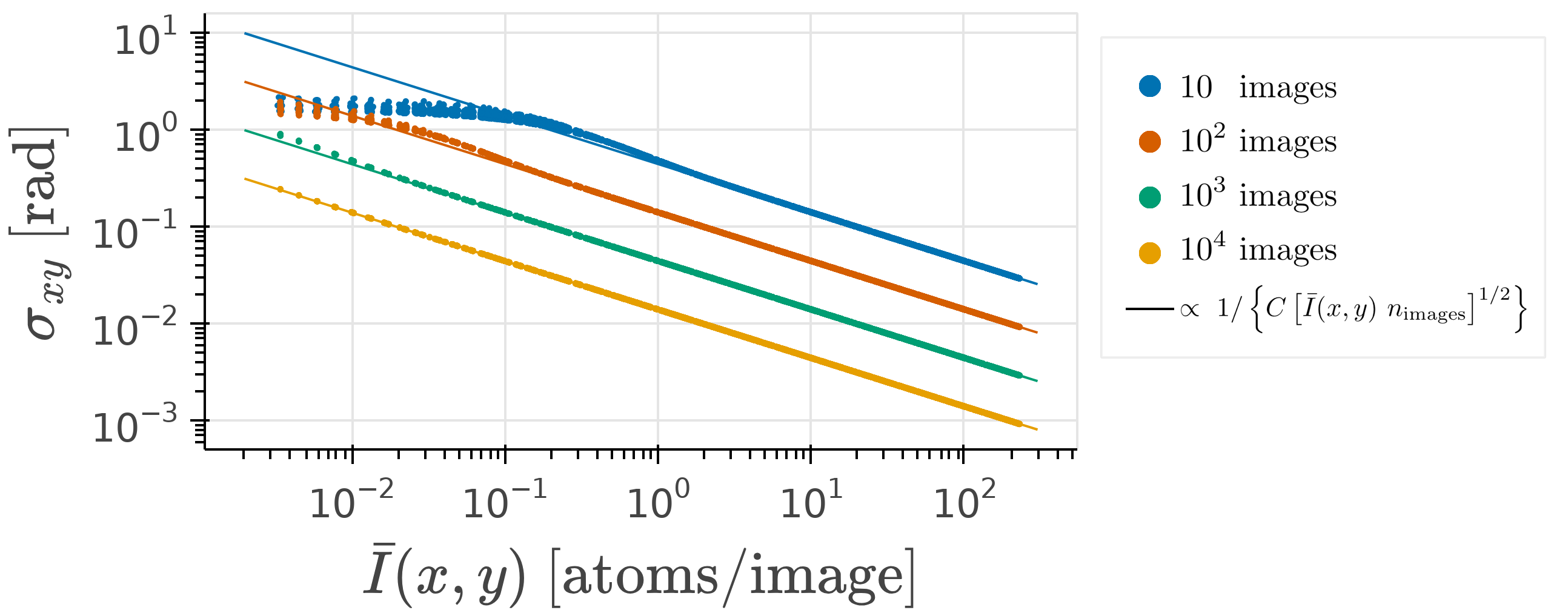}
    \caption{$\phiXYStd$ scaling with the atom number per image. This plot shows all pixels from all three phase patterns $\phiXYInp$.
    Four different values from the scan of $\nImages$ were selected and are color-coded according to the legend.
    The order in the plot from top to bottom is the same as in the legend.
    $\phiXYStd$ scales with $b/\{C \ [\nImages \IAvg]^\frac{1}{2}\}$ with $b=1.41$ as a fitted constant.
    One can see that $\phiXYStd$ of a pixel only depends on the number of atoms that populated that pixel over the course of the whole data set.
    }
    \label{fig:dPhiScaling}
\end{figure}

Similarly, $\phiXYStd$ solely scales with the total number of interfering atoms that fell into the reconstructed pixel during the image sequence as depicted in \cref{fig:dPhiScaling}.

\FloatBarrier

\section{Example Reconstruction of the Image and Spatial Phases from Experimental Data}\label{sec:expData}
In this chapter, PSPR is used to extract the spatial phase profile of two interferometry ports in a set of experimental data images.
The experimental data was taken with the apparatus described in reference~\cite{vanZoest2010}.
It uses an atom chip and laser beams for the generation and preparation of Bose-Einstein condensates (BEC).
The experimental sequence is the fountain gravimeter shown in reference~\cite{Abend2016} where the atom chip is also used as a mirror for the interrogating laser beam and therefore acts as the inertial reference of the measurement.

The atom chip has an optical coating to yield a highly reflective optical surface, but is not an ideal mirror.
Its surface has imperfections, which can lead to spatial wavefront distortions, beam intensity fluctuations and other aberrations of the employed light fields.
These can produce a spatial phase distortion in the interferometer outputs, which is examined with PSPR in this article.
A first test with a pure PCA-based evaluation on the experimental data sets already showed that the intrinsic interferometer sensitivity can be enhanced by enlarged contrast and respectively reduced technical noise~\cite{Abend2017}.

The first measurement consists of 991 images from a Mach-Zehnder interferometer operated with Bragg diffraction where each beam-splitter transfers a momentum of $\hbark{2}$ onto the atoms \cite{Giltner1995, Martin1988}, with $k$ denoting the wave number of the Bragg laser and $\hbar$ the reduced Planck's constant.
The second data set is from an interferometer with larger momentum transfer of $\hbark{6}$ via higher-order Bragg transitions and consists of 1015 images.
\Cref{fig:ExpPCAResComps} shows the results of step 2 of the PSPR algorithm (\cref{fig:ComputeGraph}).
The average intensity $\IAvg$ is clearly distributed over two ports.
This figure does not show the $\coeff{j}$ because they are unordered due to vibrations causing large phase jumps between each image.
For the first data set the first component shows the typical pattern of mainly moving atoms between the two ports, depending on the value of $\coeff{1}$.

\newcommand{\pcTableWidth}{0.2}
\begin{figure}
    \centering
    \Large

    \begin{tblr}{
        hlines,vlines,
        colspec = {Q[c]Q[c,h]Q[c,h]},
    }
      Component  & Data Set 1 $(\hbark{2})$    &   Data Set 2 $(\hbark{6})$
      \\
      $\IAvg$
      &\includegraphics[width=\pcTableWidth\textwidth]{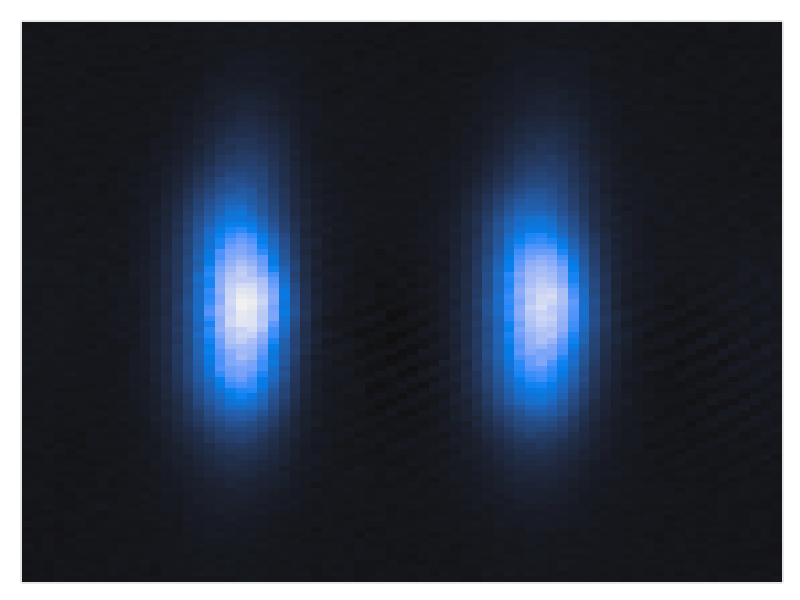}
      &\includegraphics[width=\pcTableWidth\textwidth]{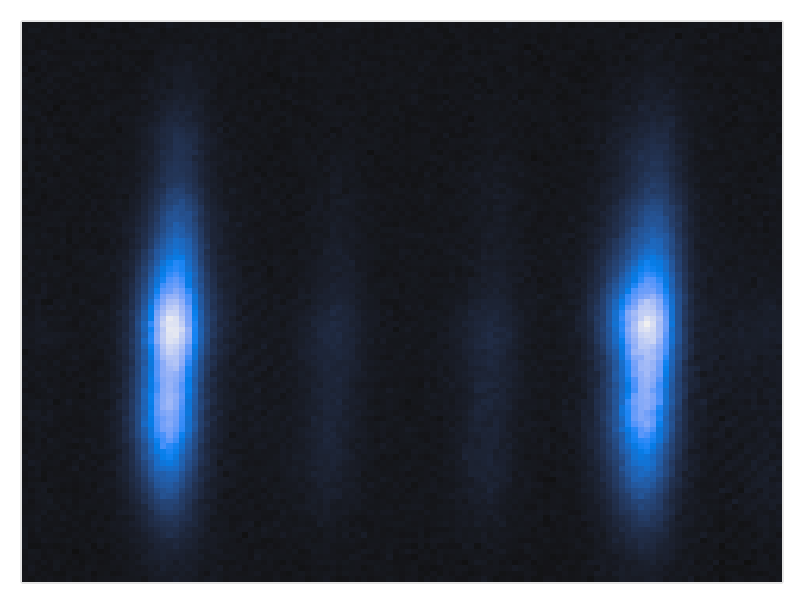}
      \\
      $\pc{1}$
      &\includegraphics[width=\pcTableWidth\textwidth]{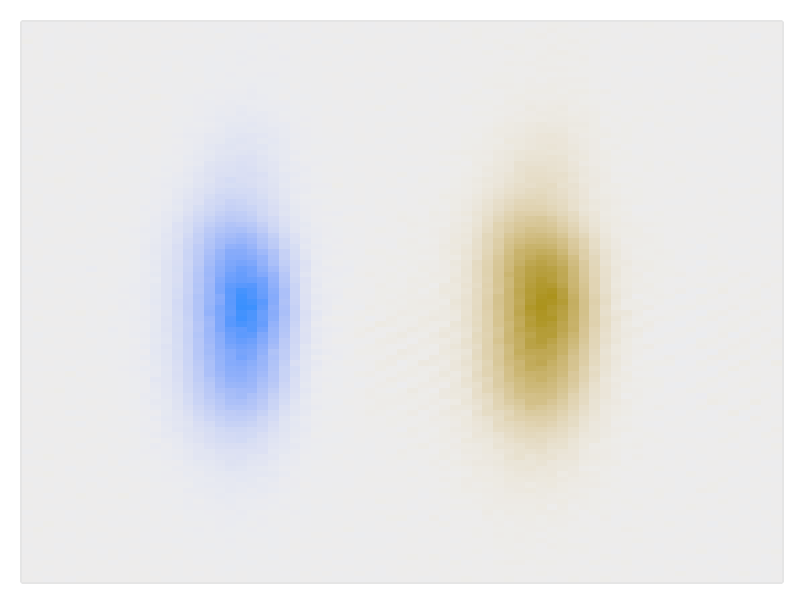}
      &\includegraphics[width=\pcTableWidth\textwidth]{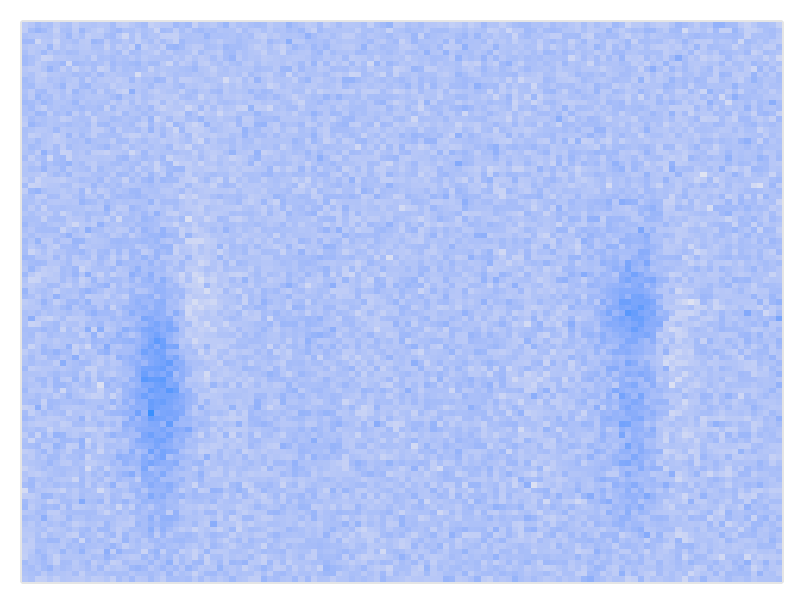}
      \\
      $\pc{2}$
      &\includegraphics[width=\pcTableWidth\textwidth]{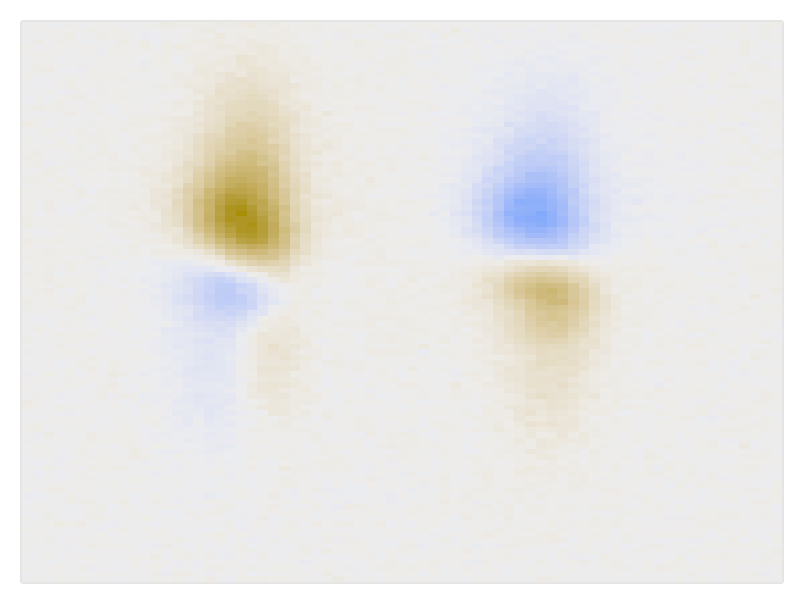}
      &\includegraphics[width=\pcTableWidth\textwidth]{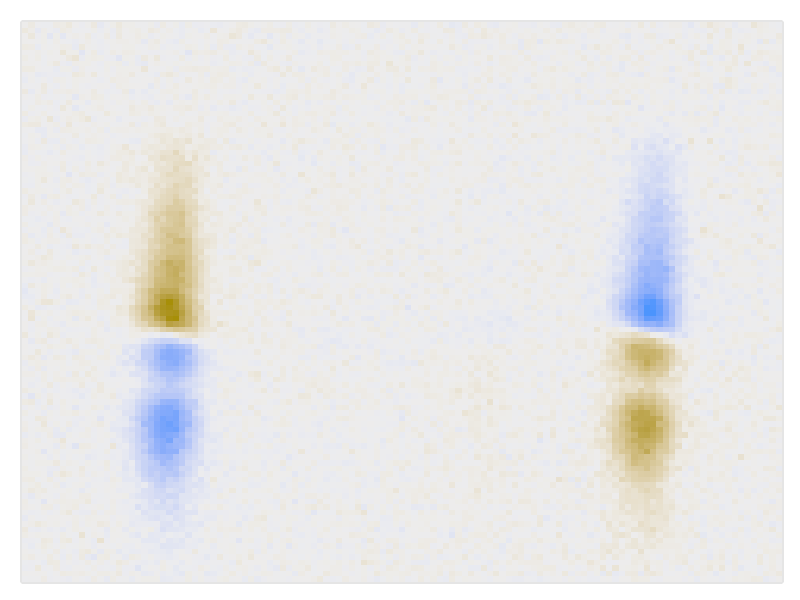}
      \\
      $\pc{3}$
      &\includegraphics[width=\pcTableWidth\textwidth]{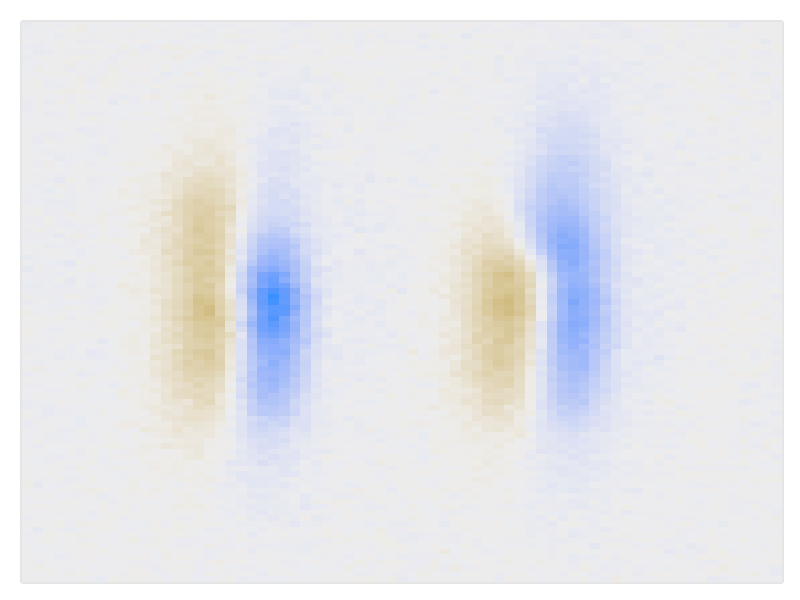}
      &\includegraphics[width=\pcTableWidth\textwidth]{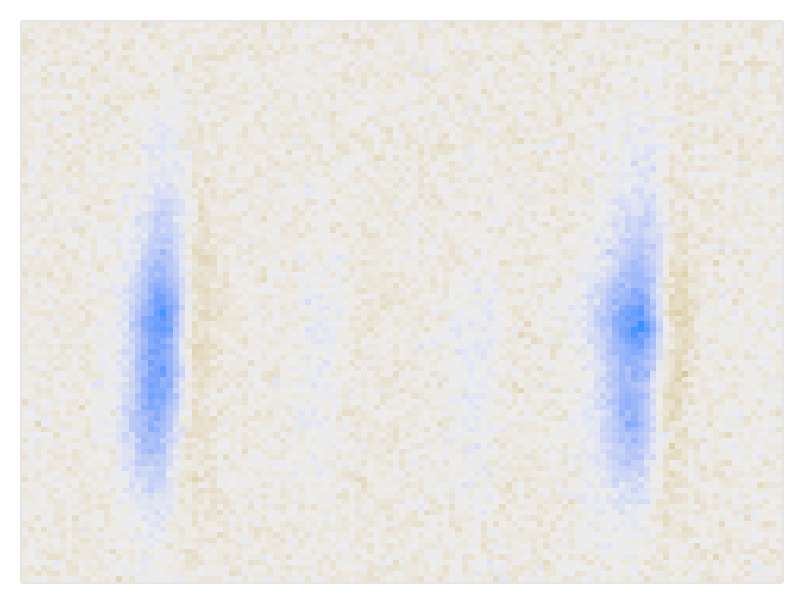}
      \\
      $\pc{4}$
      &\includegraphics[width=\pcTableWidth\textwidth]{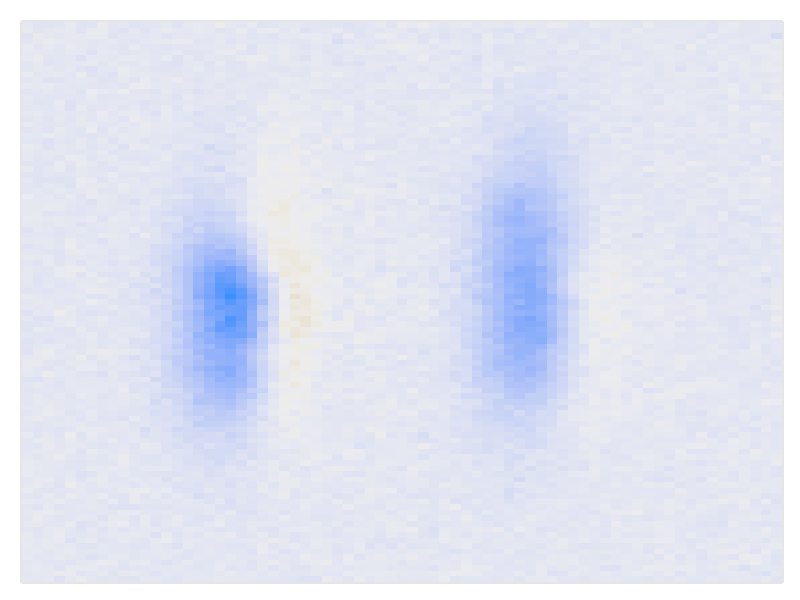}
      &\includegraphics[width=\pcTableWidth\textwidth]{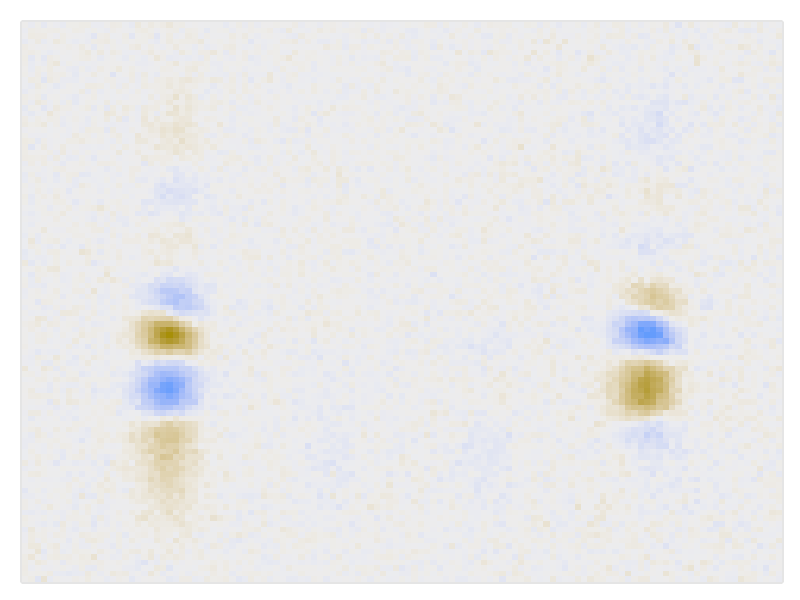}
    \end{tblr}
        \caption{
        Average intensity $\IAvg$ and the first four principal components for a Bragg diffraction based atom gravimeter featuring momentum transfers of $\hbark{2}$ (Data set 1) and $\hbark{6}$ (Data set 2). Data set 1 contains 991 images, while the results for data set 2  are reconstructed from 1015 images.
        The two ports are visible in all images.
        $\IAvg$ has a linear color map indicating the intensities while the components are plotted with a diverging color map with blue indicating negative values and yellow indicating positive values.
        $\pc{1}$ in the left column displays the main interferometer function for relatively flat phase profiles, moving atoms from left to right positive coefficients and from right to left for negative coefficients.
        The mean clearly shows the two ports, the first component is cosine-like and encodes the movement of atoms between the two ports.
        The second component is the sine-like component while the components three and four are most likely due to motion of the camera relative to the interferometer.
    }
    \label{fig:ExpPCAResComps}
\end{figure}

\begin{figure}
    \centering
    \Large

    \begin{tblr}{
        colspec = {Q[c,h]Q[c,h]},
    }
        \phantom{aaaaa}Data Set 1 $(\hbark{2})$ & \phantom{aaaaa}Data Set 2 $(\hbark{6})$\\
        \includegraphics[width=0.45\textwidth]{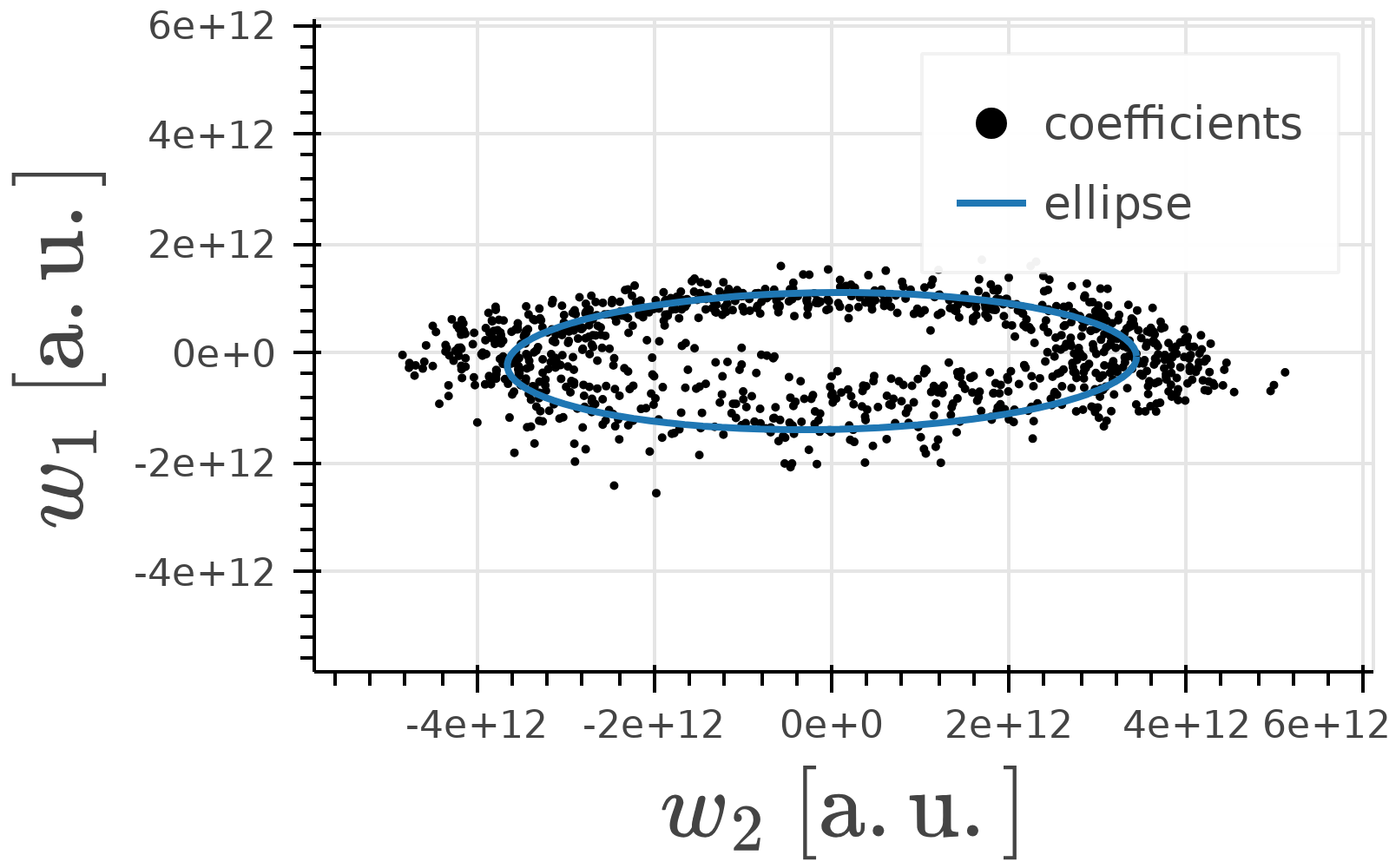}
        &\includegraphics[width=0.45\textwidth]{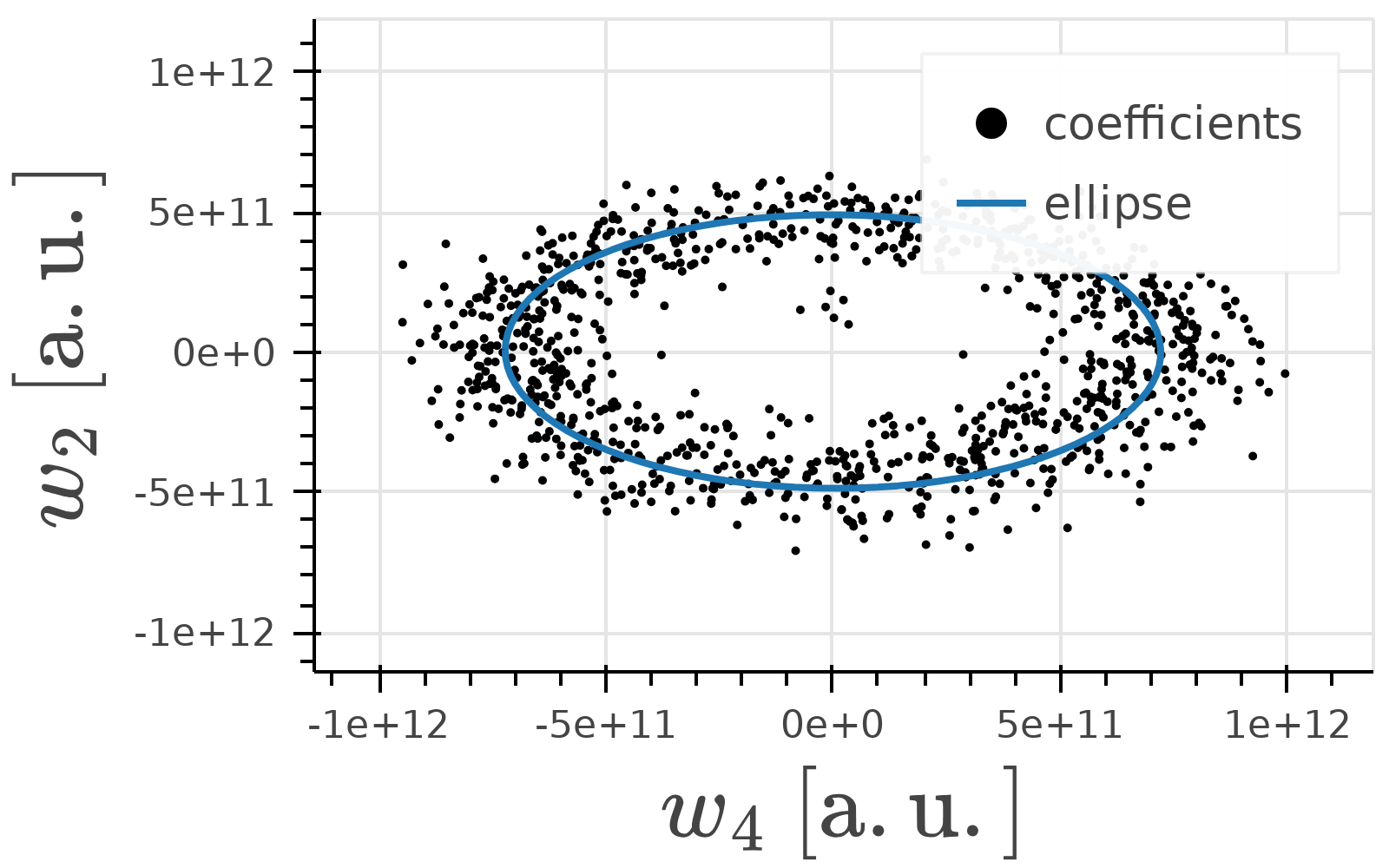}
    \end{tblr}
    \caption{
        These two plots show the correlations between the two "sine" and "cosine" components of the first and second data set (right).
        The coefficients $\coeff{j}$ are the projections of each image $\I$ onto the components $\pc{j}$ which are shown in \cref{fig:ExpPCAResComps}.
        This ellipse pattern is a signature for the sine-like and cosine-like components.
        When comparing with \cref{fig:ExpDphiRec} one can see that the stronger ellipse-shape in the first data set corresponds to a more flat spatial phase profile compared to the second data set.
        This is consistent with the simulations.
    }
    \label{fig:ExpPCAResCompCorrelation}

\end{figure}

The result of step 3 of PSPR yields the ellipses of \cref{fig:ExpPCAResCompCorrelation}.
In the second data set, other effects are stronger than the interferometry so that the $\coeff{2}$ and $\coeff{4}$ form the ellipse used for phase reconstruction.
The first data set has a weaker spatial phase profile, which causes the ellipse to be more stretched compared to the second data set with a phase profile which has a larger dynamic range of phases.

\begin{figure}
    \centering
    \Large
    \begin{tblr}{
        colspec = {Q[c,h]Q[c,h]},
    }
        Data Set 1 $(\hbark{2})$\phantom{a} & Data Set 2 $(\hbark{6})$\phantom{a}\\
        \includegraphics[width=0.45\textwidth]{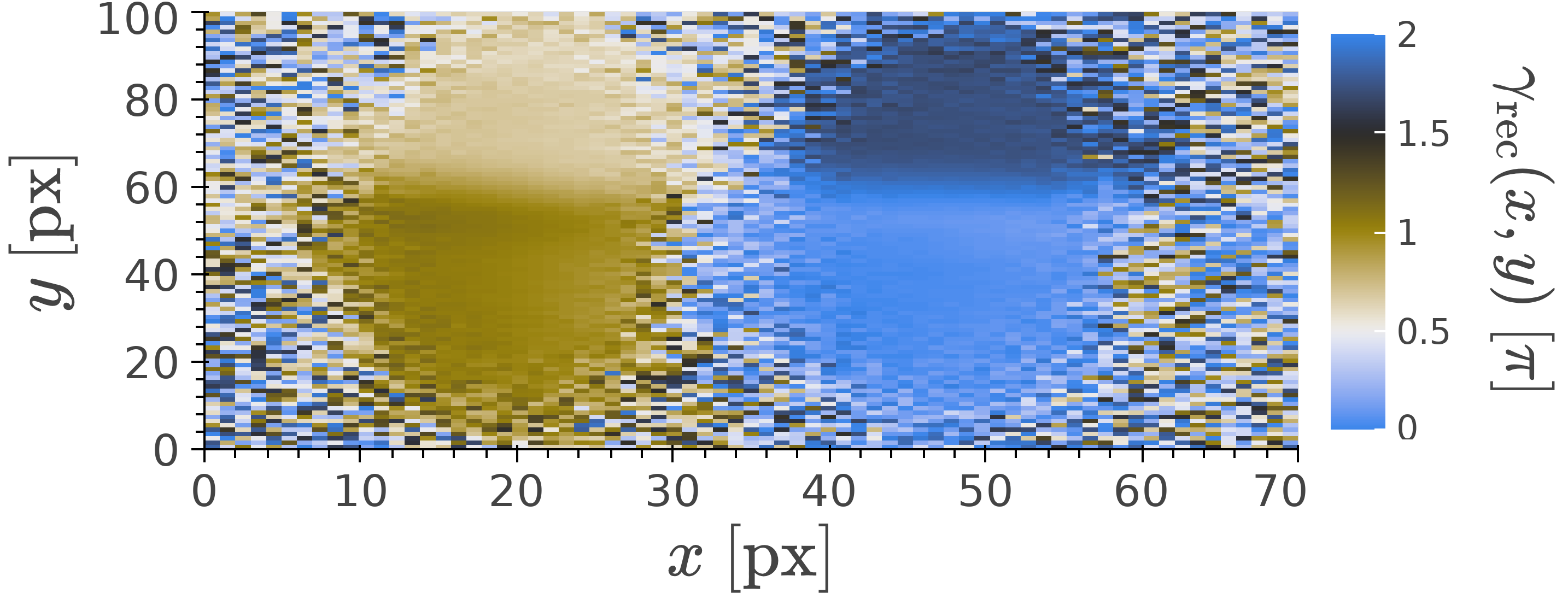}
        &\includegraphics[width=0.45\textwidth]{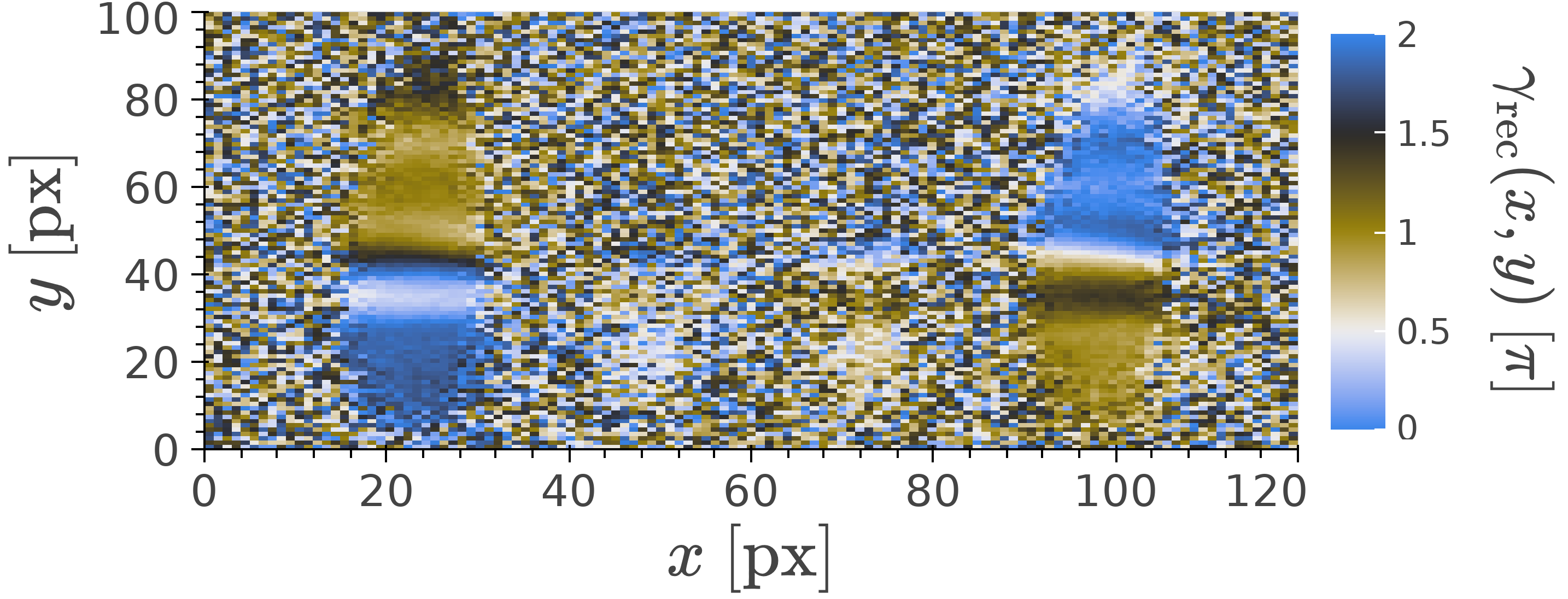}
    \end{tblr}
    \caption{
        The reconstructed spatial phase profile $\phiXYRec$ for the two experimental data sets.
        The first data set with $\hbark{6}$ momentum transfers shows much stronger distortions which is to be expected since there were far more atom-light interactions with distorted wavefronts during the sequence.
        In the areas on the sides of the ports, where there are almost no atoms, the signal strength is too weak to give a meaningful signal.
    }
    \label{fig:ExpDphiRec}
\end{figure}

The main result of this section can be seen in \cref{fig:ExpDphiRec} which shows the actual reconstruction of the spatial phase profiles when using the whole algorithm.
The second data set shows a much stronger spatial phase gradient, than the first data set, which is to be expected due to the larger number of atom-light interactions.
In the second data set with the stronger $\hbark{6}$ momentum transfer, the spatial phases are spanning a gamut of $2\pi$ so for any overall interferometer phase there are always some interfering atoms in each of the ports.
Analysing this data set without taking into account the spatial phase profile would leave a lot of useful signal unutilized and would lead to increased systematic effects.

In the second data set there are also two faint clouds between the strong clouds which are a loss channel formed by atoms that interacted at a different momentum transfer order.
These clouds indicate the possibility of utilising information from what would normally regarded as a loss channel.

\section{Conclusion and Outlook}
\label{sec:Discussion}
The PSPR algorithm developed in this article allows to reconstruct the spatial phase profiles in image sets of atom interferometry ports.
Its performance was proven to theoretically reach the machine precision level.
The algorithm was then successfully applied to reconstruct the experimental phase front pattern seen in the ports of an atom gravimeter.
These results point towards a future use of PSPR in the analysis of atom interferometers embracing and utilizing non-plane wavefronts to make measurements more accurate and versatile instead of simply smoothing the inevitable wavefronts of the interrogation lasers.

It is a first step towards measuring wavefront aberrations using the exact atom-light phase profiles at the positions where the interaction occurs during experimental sequences.
This will improve the understanding of the systematics in matter wave systems by varying different parameters in the setup and characterizing the impact on the spatial phase map.

All parts of PSPR can be implemented very efficiently even on hardware with limited performance and memory.
For PCA, randomized out-of-core algorithms exist, which can achieve the decomposition without ever loading the whole data set into the main memory \cite{Halko2011}.
Furthermore, the ellipse fitting algorithm is stable even with in the presence of significant instrument noise levels and completely iteration-free \cite{Halir1998a}.

PSPR performs a lossy compression that is well adapted to interferograms of atom interferometers.
It exploits that the spatial phase imprints do not vary much during each measurement while still allowing arbitrary phase patterns which can be tracked over the long-term operation of the interferometer.
Using this knowledge it reduces the three-dimensional $\I$ into the much smaller one-dimension $\phiI$ and two-dimensional $\IAvg$ and $\phiXY$.

Additionally, PSPR offers two heuristics that can be used to automatically detect when the atom interferometer deviates too much from its expected behavior.
The first heuristic is how well the two components that are used for the phase reconstruction describe the whole image set.
This can be computed, for example, by projecting the images into that basis and evaluating their deviation from the raw data.
The second heuristic is how well the ellipse fits onto the coefficients giving a clear indicator for how accurately the two selected components actually describe the expected interferogram defined in \cref{eq:imgModel}.

Having the potential to realize the phase extraction and tracking of key performance characteristics directly on the specific atom interferometry device, PSPR helps to make autonomous operation possible.
For instance, as part of the setup of the interferometer one can calibrate expectation thresholds for how well the two used PCA components and the ellipse fit explain the data.
When these thresholds are violated, the device may, depending on the use case, launch debugging routines, safeguards, warnings or special calibration routines to mitigate the problems or explicitly request user intervention.

For future atom interferometers deployed in real-world environments on ground and in space \cite{Geiger2020a, Ahlers2022, Leveque2023}, it will be necessary to fully automate the data analysis directly on an embedded processor.
For field applications, the number of measurements will be too large to be analyzed with manual intervention by non-experts.
Moreover, the storage requirements for the raw data also become a limiting factor.
Finally, in space or in environments where communication is denied, bandwidth to send raw data to be processed is not available or insufficient.
For all these reasons, PSPR will be a well suited on-board tool.

\begin{acknowledgments}
  We thank Klemens Hammerer for his insights about shot noise models and Víctor J. Martínez-Lahuerta and Matthew Glaysher for carefully reading the manuscript.
  Additionally we thank Christian Struckmann for helpful discussions about phase differences and uniqueness.
  This work was funded by the Deutsche Forschungsgemeinschaft (German Research Foundation) under Germany’s Excellence Strategy (EXC-2123 QuantumFrontiers Grants No. 390837967), through CRC 1227 (DQ-mat) within Projects No. A05 and through the QuantERA 2021 co-funded project No. 499225223 (SQUEIS), and the German Space Agency (DLR) with funds provided by the German Federal Ministry of Economic Affairs and Energy (German Federal Ministry of Education and Research (BMBF)) due to an enactment of the German Bundestag under Grants No. DLR 50WM1952 (QUANTUS-V-Fallturm), No. 50WM2250A (QUANTUS plus), No. 50WM2450A (QUANTUS-VI), No. 50WM2245A (CAL-II), No. 50WM2263A (CARIOQA-GE), No. 50WM2253A (AI-Quadrat), and No. 50NA2106 (QGYRO+).
\end{acknowledgments}

\appendix*
\section{Computing the Minimal Difference Between Two Phase Results}

The phase offset degree of freedom $c$ can be determined using $\angleDiff$:
\begin{align*}
    \PhiOffsetDiff &= \angleDiff[\phiIInp, \phiIRec]\\
    x &= \mathrm{mean}\{\cos[\PhiOffsetDiff]\}\\
    y &= \mathrm{mean}\{\sin[\PhiOffsetDiff]\}\\
    c &= \mathrm{arctan2}(y,x).
\end{align*}
The value of $c$ is the average of all differences between $\phiIInp$ and $\phiIRec$ while not making averaging errors due to the phase wrap-around after $2\pi$.
That average is the angle of the average position of the points of all angles on the unit circle.

The sign $s$ is determined by computing $c$ for both $\angleDiff[\phiIInp, \phiIRec]$ and $\angleDiff[\phiIInp, -\phiIRec]$ and choosing the sign for which that difference is smaller.

\bibliography{bibliography, software}

\end{document}